\documentclass[sort&compress,12pt]{elsarticle}

\usepackage{graphicx}
\usepackage{amsmath}

\usepackage{natbib}
\usepackage{amssymb}

\usepackage{amsthm}

\usepackage{lineno}
\usepackage{subfig}
\usepackage{enumerate}
\usepackage{fullpage}
\usepackage{algorithm}
\usepackage{algpseudocode}
\usepackage{xcolor}
\usepackage[colorinlistoftodos]{todonotes}

\biboptions{sort,compress} 

\usepackage{xcolor}

\newcommand*\patchAmsMathEnvironmentForLineno[1]{%
	\expandafter\let\csname old#1\expandafter\endcsname\csname #1\endcsname
	\expandafter\let\csname oldend#1\expandafter\endcsname\csname end#1\endcsname
	\renewenvironment{#1}%
	{\linenomath\csname old#1\endcsname}%
	{\csname oldend#1\endcsname\endlinenomath}}%
\newcommand*\patchBothAmsMathEnvironmentsForLineno[1]{%
	\patchAmsMathEnvironmentForLineno{#1}%
	\patchAmsMathEnvironmentForLineno{#1*}}%
\AtBeginDocument{%
	\patchBothAmsMathEnvironmentsForLineno{equation}%
	\patchBothAmsMathEnvironmentsForLineno{align}%
	\patchBothAmsMathEnvironmentsForLineno{flalign}%
	\patchBothAmsMathEnvironmentsForLineno{alignat}%
	\patchBothAmsMathEnvironmentsForLineno{gather}%
	\patchBothAmsMathEnvironmentsForLineno{multline}%
}

\usepackage{graphicx}
\usepackage{amssymb}
\usepackage{amsthm}
\usepackage{bbm}
\usepackage{lineno}
\usepackage{fullpage}
\usepackage[colorinlistoftodos]{todonotes}
\usepackage{url}
\usepackage{listings}

\usepackage{color,soul}
\usepackage{enumitem}
\usepackage{mathtools}

\definecolor{lightblue}{rgb}{.90,.95,1}
\definecolor{darkgreen}{rgb}{0,.5,0.5}

\definecolor{lightgreen}{rgb}{.90,1,0.90}

\newcommand{\bstau}{\boldsymbol{\tau}}

\newcommand{\bs}[1]{\boldsymbol{#1}}

\usepackage{array}
\usepackage{multirow}
\graphicspath{ {./figs/} }
\newcolumntype{P}[1]{>{\centering\arraybackslash}m{#1}}
\newcolumntype{L}[1]{>{\raggedright\let\newline\\\arraybackslash\hspace{0pt}}m{#1}}
\newcolumntype{C}[1]{>{\centering\let\newline\\\arraybackslash\hspace{0pt}}m{#1}}
\newcolumntype{R}[1]{>{\raggedleft\let\newline\\\arraybackslash\hspace{0pt}}m{#1}}

\graphicspath{ {./figs/} }
\journal{Physical Review Fluids}

\begin{document}

\begin{frontmatter}



 \title{A Comprehensive Physics-Informed Machine Learning Framework for Predictive Turbulence Modeling\tnoteref{jref}}



\tnotetext[jref]{This manuscript is unpublished. The work presented herein has been significantly improved and published separately as:
J.-L. Wu, H. Xiao and E. G. Paterson. \textbf{Physics-informed machine learning approach for augmenting turbulence models: A comprehensive framework.} \emph{Physical Review Fluids}. 3, 074602, 2018. DOI: 10.1103/PhysRevFluids.3.074602}

\author[vt]{Jian-Xun Wang}
\author[vt]{Jinlong Wu}
\author[snl]{Julia Ling}
\author[jops]{Gianluca Iaccarino}
\author[vt]{Heng Xiao\corref{corxh}}
\cortext[corxh]{Corresponding author. Tel: +1 540 231 0926}
\ead{hengxiao@vt.edu}
\address[vt]{Department of Aerospace and Ocean Engineering, Virginia Tech, Blacksburg, VA}
\address[snl]{Sandia National Laboratories, Livermore, CA}
\address[jops]{Department of Mechanical Engineering, Stanford University, Stanford, CA}

\begin{abstract}
Although an increased availability of computational resources
has enabled high-fidelity simulations (e.g., large eddy
simulations and direct numerical simulation) of turbulent flows, the Reynolds-Averaged 
Navier--Stokes (RANS) models are still the dominant tools for industrial 
applications. However, the predictive capabilities of RANS models
are limited by potential inaccuracy driven by hypotheses in the
Reynolds stress closure. Recently, a Physics-Informed Machine Learning (PIML) 
approach has been proposed to learn the functional form of Reynolds 
stress discrepancy in RANS simulations based on available data. 
It has been demonstrated that the learned discrepancy function can be 
used to improve Reynolds stresses in different flows where data are not available. 
However, owing to a number of challenges, the improvements have been
demonstrated only in the Reynolds stress prediction but not in the corresponding propagated 
quantities of interest (e.g., velocity field), which is still an \emph{a priori} study. 
In this work, we introduce and demonstrate the procedures toward a complete
PIML framework for predictive turbulence modeling, including learning Reynolds stress 
discrepancy function, predicting Reynolds stresses in different flows, and propagating the 
predicted Reynolds stresses to mean flow fields. The process of Reynolds 
stress propagation and predictive accuracy of the propagated velocity field 
are investigated. To improve the learning-prediction performance, the input
features are enriched based on an integrity basis of invariants. The fully developed 
turbulent flow in a square duct is used as the test case. The discrepancy model
is trained on flow fields obtained from several Reynolds numbers and evaluated
on a duct flow at a Reynolds number higher than any of the training cases.
The predicted Reynolds stresses are propagated to velocity field through RANS equations. 
Numerical results show excellent predictive performances in both Reynolds stresses and 
their propagated velocities, demonstrating the merits of the PIML approach in 
predictive turbulence modeling.
\end{abstract}

\begin{keyword}
  machine learning \sep turbulence modeling\sep Reynolds-Averaged Navier-Stokes
  equations \sep data-driven approach
\end{keyword}
\end{frontmatter}

\section{Introduction}
\label{sec:intro}
Computational fluid dynamic (CFD) has been widely used to simulate turbulent
flows. Although the rapidly increasing availability of computational resource enables 
high-fidelity simulations, e.g., Large Eddy Simulation (LES) and Direct Numerical 
Simulation (DNS), it is not yet computationally feasible to routinely apply them for complex,
industrial flows. Therefore, Reynolds-Averaged Navier-Stokes (RANS) models, 
where empirical closures are used to model the Reynolds stresses, are still the dominant 
tools for practical engineering problems. However, RANS predictions are known to be 
unreliable for flows with strong pressure gradients, curvature, or separation~\cite{craft1996development}. 
This is because some assumptions (e.g., the Boussinesq assumption) made in the closure 
models are not universally valid, and thus model-form errors are potentially introduced in predicting
the regime-dependent, physics-rich phenomena of turbulent flows. These assumptions are typically in 
the form of functional dependency between mean flow properties and turbulent quantities. 
They have been traditionally formulated by using intuition, physical observations, and theoretical 
constraints. Although advanced RANS models~\cite[e.g.,][]{launder1975progress,wallin2000explicit} 
have been developed in the past decades, a universally applicable one is still lacking. 
The traditional process of RANS model development solely based on physical understanding 
and reasoning seems to be insufficient to significantly improve the predictive capability.

Recently, the increasing availability of high-fidelity data sets from both DNS
and experiments makes it possible to use the data-driven approach as a complement 
to the physics-based approach to improve the predictive capability of RANS models.
The past several years have witnessed a few efforts to develop data-driven turbulence 
modeling approaches by using machine learning algorithms~\cite[e.g.,][]{milano2002neural,dow11quanti,
tracey2013application,parish2016paradigm,ling2015evaluation,mfu1,Wang2016}. 
Generally, machine learning refers to a process of using data to build explicit or implicit 
functions of responses with respect to input variables (features). The trained functions 
can be evaluated to predict cases where data are not available. In the context of 
turbulence modeling, different machine learning approaches aim to achieve a similar overarching 
goal, i.e., improving the predictive capability of turbulence models. However, so far there 
is no consensus yet on the choices of learning responses and input features to better achieve this 
goal. Dow and Wang~\cite{dow11quanti} chose the discrepancy field $\Delta\nu_t$ in
turbulent viscosity as the response, while Duraisamy and co-workers~\cite{parish2016paradigm,
singh16using,singh2016machine} introduced a full-field multiplicative discrepancy 
factor~$\beta$ into the production term of the transport equation as the learning target. 
Although both the inferred $\Delta\nu$ and $\beta$ are demonstrated to be able to extrapolate 
to certain extents, they are still modeled quantities and have less physical interpretations. 
Xiao et al.~\cite{mfu1} directly inferred the discrepancies in RANS simulated Reynolds stresses by 
using sparse velocity measurements. Wu et al.~\cite{mfu3} further demonstrated these inferred 
Reynolds stress discrepancies can be extrapolated to closely related flows. Although the response 
chosen in~\cite{mfu1,mfu3} is a physical quantity, an intrinsic limitation lies on their choice of 
input features (i.e., physic coordinates $\mathbf{x}$). As a result, the prediction can only be 
applied to the flows in the same geometry at the same location. Duraisamy and 
co-workers~\cite{tracey2013application,duraisamy2015new} used non-dimensional flow and 
model variables to construct the input feature space for calibrations of low-fidelity models. 
However, their input feature space was constructed with a very small number (three) of features 
and the invariant property was not fully considered. Ling et al.~\cite{ling2016machine} pointed 
out the merits of embedding the invariance properties into machine learning process. They explored 
several machine learning models for classifying the regions where RANS assumptions would 
fail~\cite{ling2015evaluation}. Most recently, they also attempted to directly predict the anisotropy 
tensors of the Reynolds stresses by using random forests~\cite{ling2016uncertainty} and deep neural networks~\cite{ling2016reynolds}. 

By comprehensively considering the physical interpretability of learning targets and the invariance 
property of input features, Wang et al.~\cite{Wang2016} proposed a physics-informed machine 
learning (PIML) approach to learn the functional forms of Reynolds stress discrepancy on 
its six physically meaningful components (i.e., magnitude, shape, and orientation of Reynolds 
stress tensor) with respect to a group of ten invariant mean flow features. They successfully
demonstrated that the trained discrepancy model can be used to improve the RANS-modeled Reynolds 
stresses in flows with different configurations. However, this work is still \emph{a priori} investigation,
since the improvements are demonstrated only in Reynolds stresses but not in their propagated velocities. 
There are a number of challenges associated with propagating forward the corrected Reynolds stresses 
through RANS equations to obtain the mean velocity and pressure fields. For example, the high-fidelity 
data themselves used for training must be highly accurate to obtain a precise mean velocity field 
after propagation. Moreover, the machine learning model should improve predictions of the mean flow variables, 
which requires not only the pointwise Reynolds stress predictions but also their derivatives to be improved, 
since it is the divergence of Reynolds stress that appears in the RANS momentum equations. The objective of 
this work is to introduce the procedures toward a complete PIML framework and demonstrate its capability
of improving both the Reynolds stresses and their propagated mean velocities in a relatively less 
challenging scenario with reliable training data. To improve the predictive accuracy of the machine learning 
model, the input feature space adopted in~\cite{Wang2016} is expanded by constructing an integrity 
basis of invariants of mean flow variables. The systematic approach of invariant feature construction 
proposed in~\cite{ling2016machine} is employed to expand the input space for given raw tensorial 
mean flow variables. The fully developed turbulent flow in a square duct is used to demonstrate the
merits of the proposed method. The discrepancy model is trained on duct flows from several 
Reynolds numbers and evaluated on a duct flow at a Reynolds number higher than any of the training 
cases. The predicted Reynolds stresses are propagated to the mean velocity field through the RANS 
equations, and the accuracy of the propagated mean velocities are investigated and the difficulties 
associated with the propagation are discussed. Although current work is developed in the context of 
turbulence modeling, it has potential implications in many other fields in which the governing equations 
are well understood but empirical closure models are used for the unresolved physical process.

%

The rest of the paper is organized as follows. Section~\ref{sec:meth} introduces building 
blocks of the Physics-Informed Machine Learning (PIML) framework, including the construction of
input feature space, representation of Reynolds stress discrepancy as the response, 
construction of regression function of the discrepancy with respect to input features, and 
propagation of corrected Reynolds stresses to mean velocity field. Section~\ref{sec:result} 
presents the numerical results to demonstrate the prediction performance of the proposed 
framework and merits of systematical expansion of input space. The concept of 
``physics-informed machine learning'' and future perspectives of the current framework 
are discussed in Section~\ref{sec:discussion}. Finally, Section~\ref{sec:conclusion} concludes the paper.

\section{Methodology}
\label{sec:meth}
In this section, the Physics-Informed Machine Learning (PIML) framework for predictive turbulence 
modeling is summarized. Its key procedures and components, including construction of the input 
feature set, choice of output responses, and building of regression functions, are discussed. 

\subsection{Overview of PIML Framework}
 \label{sec:meth:overview}
The aim of the present work is to introduce and demonstrate the PIML framework for predictive 
turbulence modeling. Specifically, given high-fidelity data (e.g., Reynolds stresses from DNS 
simulations) from a set of training flows, the framework aims to improve the standard RANS 
prediction for different flows for which DNS data are not available. Here, \emph{training flow} 
refers to the flow with high-fidelity data, which are used to train the machine learning model. Accordingly, 
\emph{test flow} (prediction flow) is the flow to be predicted. Generally, training flows should share 
similar flow physics with the test flow, so that the model does not have to extrapolate. 
This scenario is typical in the engineering design process, where data are available for 
some flows and predictions are required for different flows with slightly changed configurations 
(e.g., different Reynolds numbers or slightly changed geometries) but without data. 

\begin{figure}[htbp]
	\centering
	\includegraphics[width=0.9\textwidth]{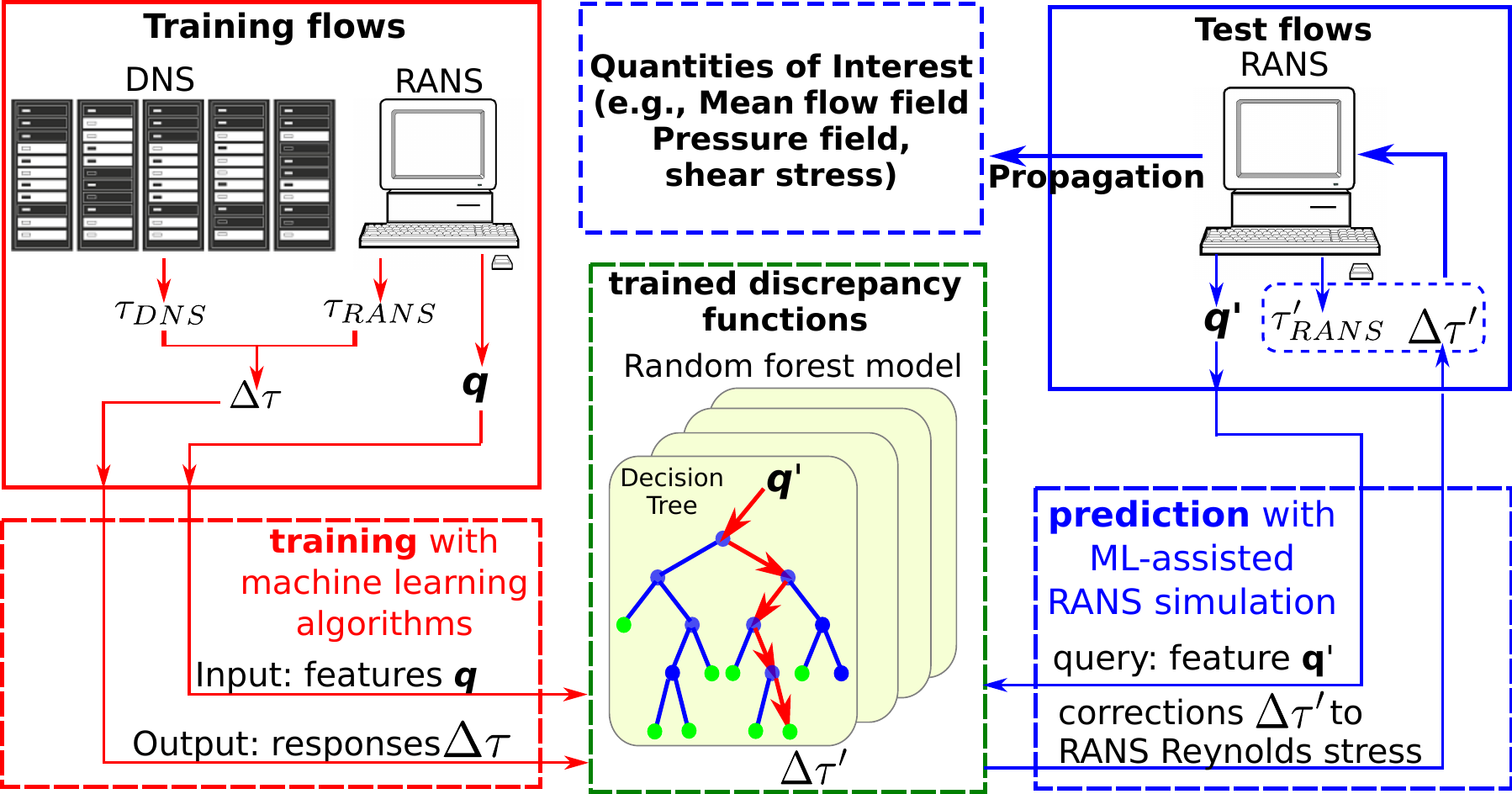}
	\caption{Schematic of Physics-Informed Machine Learning (PIML) framework
	for predictive turbulence modeling. Conduct both DNS and RANS simulations 
	for training flows to obtain the data (i.e., mean flow features $\mathbf{q}$ as the 
	input and Reynolds stress discrepancies $\Delta\bs{\tau}$ as the output). 
	The training data are then used to train the random forest models of discrepancy functions. 
	To perform the prediction, RANS simulations are conducted for the test flow to obtain 
	the mean flow features, which can be used to query the trained discrepancy functions 
	and then correct the corresponding RANS-modeled Reynolds stresses. The corrected
	Reynolds stresses are propagated through RANS equations to the QoIs (e.g. mean velocity field)}
	\label{fig:piml}
\end{figure}
In RANS simulations model-form uncertainties stem from the RANS-modeled Reynolds stresses. 
Therefore, the aim of the machine learning is to extract the functional form of discrepancies in 
RANS-modeled Reynolds stresses from data. Figure~\ref{fig:piml} shows a schematic of the 
PIML framework. The overall procedure can be summarized as follows:
 \begin{enumerate}
 	\item Perform baseline RANS simulations on both the training flows and the test flow.
 	\item Compute the input feature field $\mathbf{q}(\mathbf{x})$ based on the local 
 	RANS flow variables. 
 	\item Compute the discrepancies field $\Delta \bs\tau(\mathbf{x})$ in the RANS-modeled 
 	Reynolds stresses for the training flows based on the high-fidelity data.
 	\item Construct regression functions $ f: \mathbf{q} \mapsto \Delta \bs\tau$ for the
 	discrepancies based on the training data prepared in Step 3, using machine learning algorithms.
 	\item Compute the Reynolds stress discrepancies for the test flow by querying the regression
 	functions. The Reynolds stresses can subsequently be obtained by correcting the baseline RANS 
 	predictions with the evaluated discrepancies.
 	\item Propagate the corrected Reynolds stresses to the mean velocity field by solving the RANS 
 	equations with the corrected Reynolds stress field.
 \end{enumerate}
 There are four essential components in the PIML framework: (1) construction of the input feature set,
 (2) representation of the Reynolds stress discrepancy as the response, (3) construction of 
 the regression function of the discrepancy with respect to input features, and (4) propagation of
 corrected Reynolds stresses to mean velocities. The details of the framework are introduced
 below, and the procedures to systematically construct and expand the input features are 
 highlighted. 
 
 \subsection{Construction of Invariant Input Feature Set}
 \label{sec:meth:input}
Constructing a reasonable input feature space is of pivotal importance to the performance of
machine learning models. First, the input features should be rich enough to differentiate data 
points in the feature space and better describe the functional relation between inputs and 
responses. Moreover, it is more desirable to embed known invariance properties into the construction 
process to achieve improved generalization. Wang et al.~\cite{Wang2016} employed a 
set of ten invariant features to build random forest regressors. Formulation of these 
invariant features heavily relied on physical understanding and reasoning. However, these 
ten features are not necessarily rich enough to represent all possible polynomial invariants of 
the local mean flow variables. Therefore, a systematic methodology of constructing a complete
invariant input set from a group of given tensorial variables as suggested by Ling et al.~\cite{ling2016machine} 
is employed in the current work. Specifically, given a finite collection of raw mean flow variables 
(i.e., tensors or vectors), a finite integrity basis of invariants can be constructed. Any 
scalar invariant function of the raw inputs can be formulated as a function of the corresponding 
invariant basis. 

The first step is to identify the raw tensors and vectors. When choosing these raw input materials, 
they should be able to represent physical characteristics of the mean flow. Generally, these raw 
variables can be chosen in the same way as the traditional turbulence modeler does for developing 
advanced turbulence models. On the basis of the components used in the conventional turbulence 
modeling, a set of four raw inputs are identified as
\begin{equation}
\label{eq:raw}
\mathcal{Q} =  \{\mathbf{S}, \bs{\Omega}, \nabla p, \nabla k \},
\end{equation} 	 
where $\mathbf{S}$ and $\bs{\Omega}$ are strain rate and rotation rate tensors, respectively;
$\nabla p$ and $\nabla k$ are the gradients of pressure and turbulence kinetic energy (TKE), 
respectively. The four raw tensors and vectors above are assumed to represent the important 
physical characteristics of the mean flow, which are also widely used as crucial gradients in 
traditional turbulence modeling. For example, the combinations of $\mathbf{S}$ and 
$\bs{\Omega}$ were used to construct nonlinear eddy viscosity models~\cite{pope2001turbulent}. 
\begin{table}[htbp] 
	\centering
	\caption{
		Non-dimensional raw mean flow variables used to construct the invariant basis.  
		The normalized feature $\hat{\alpha}$ is obtained by normalizing the corresponding 
		raw input $\alpha$ with normalization factor $\beta$ according to 
		$\hat{\alpha} = \alpha / (|\alpha| + |\beta|)$. Notations are as follows: $\mathbf{U}$ is 
		mean velocity vector, $k$ is turbulence kinetic energy (TKE), $\rho$ is fluid density, 
		$\varepsilon$ is the turbulence dissipation rate, $\mathbf{S}$ is the strain rate tensor, 
		$\bs{\Omega}$ is the rotation rate tensor, $\| \cdot \|$ indicate matrix norm.  }
	\label{tab:featureRaw}
	\begin{tabular}{P{2.5cm} | P{3cm}  P{3.0cm}  P{5.0cm} }	
		\hline
		Normalized raw input $\hat{\alpha}$  & description & raw input $\alpha$ &
		normalization factor $\beta$  \\ 
		\hline
		$\hat{\mathbf{S}}$  & strain rate tensor&
		$\mathbf{S}$ & 
		$\dfrac{\varepsilon}{k}$\\  
		\hline
		$\hat{\bs{\Omega}}$  & rotation rate tensor & $\bs{\Omega}$ &
		$\|\mathbf{\Omega}\|$\\ 
		\hline
		$\hat{\nabla p}$  & Pressure gradient &
		$\nabla p$ & $\rho\|\mathbf{U} \cdot \nabla\mathbf{U}\|$\\
		\hline
		$\hat{\nabla k}$  & Gradient of TKE& $\nabla k$ & $\dfrac{\varepsilon}{\sqrt{k}}$ \\ 
		\hline					 								
	\end{tabular}
\end{table}
To ensure non-dimensionality of the raw inputs, the normalization scheme used 
in~\cite{ling2015evaluation} is adopted. Each element $\alpha$ in the raw input 
set~$\mathcal{Q}$ is normalized by a corresponding normalization factor $\beta$ 
as  $\hat{\alpha} = \alpha / (|\alpha| + |\beta|)$. Table~\ref{tab:featureRaw} shows 
all normalization factors of the raw input variables. Based on Hilbert basis 
theorem~\cite{johnson2016handbook}, a finite integrity basis of invariants 
for this set $\hat{\mathcal{Q}}$ of normalized raw inputs can be constructed.
\begin{table}[htbp]  
	\centering
	\caption{Minimal integrity bases for symmetric tensor $\hat{\mathbf{S}}$ and antisymmetric tensors $\hat{\bs{\Omega}}$, $\hat{\mathbf{A}}_{p}$, and $\hat{\mathbf{A}}_{k}$. In the implementation, $\hat{\mathbf{S}}$ is the rate of strain tensor, $\hat{\bs{\Omega}}$ is the rate of rotation tensor; $\hat{\mathbf{A}}_{p}$ and $\hat{\mathbf{A}}_{k}$ are the antisymmetric tensors associated with 
	pressure gradient $\nabla \hat{P}$ and the gradient of turbulent kinetic energy $\nabla \hat{k}$;
	$n_S$ and $n_A$ denote the numbers of symmetric and antisymmetric raw tensors for the bases;
	an asterisk ($*$) on a term means to include all terms formed by cyclic permutation of labels of anti-symmetric tensors. Note the invariant bases are traces of the tensors in the third column. 
	}
	\label{tab:basis}
	\begin{tabular}{c|C{2.5cm}|C{9.5cm}}	
		\hline
		$(n_S, n_A)$ &  feature index &  invariant bases$^{(\mathrm{a})}$\\
		\hline
		(1, 0) & 1--2 & $\hat{\mathbf{S}}^2$, $\hat{\mathbf{S}}^3$ \\
		\hline
		(0, 1)& 3--5 & $\hat{\bs{\Omega}}^2$,  $\hat{\mathbf{A}}_{p}^2$,  $\hat{\mathbf{A}}_{k}^2$ \\
		\hline
		\multirow{3}{*}{(1, 1)} & \multirow{3}{*}{6--14} 
		& $\hat{\bs{\Omega}}^2 \hat{\mathbf{S}}$, $\hat{\bs{\Omega}}^2 \hat{\mathbf{S}}^2$, $\hat{\bs{\Omega}}^2 \hat{\mathbf{S}} \hat{\bs{\Omega}} \hat{\mathbf{S}}^2$;\\
		&& $\hat{\mathbf{A}}_{p}^2 \hat{\mathbf{S}}$, $\hat{\mathbf{A}}_{p}^2 \hat{\mathbf{S}}^2$, $\hat{\mathbf{A}}_{p}^2 \hat{\mathbf{S}} \hat{\mathbf{A}}_{p} \hat{\mathbf{S}}^2$;\\
		&& $\hat{\mathbf{A}}_{k}^2 \hat{\mathbf{S}}$, $\hat{\mathbf{A}}_{k}^2 \hat{\mathbf{S}}^2$, $\hat{\mathbf{A}}_{k}^2 \hat{\mathbf{S}} \hat{\mathbf{A}}_{k} \hat{\mathbf{S}}^2$; \\
		\hline
		(0, 2)& 15--17 & $\hat{\bs{\Omega}} \hat{\mathbf{A}}_{p}$, $\hat{\mathbf{A}}_{p} \hat{\mathbf{A}}_{k}$, $\hat{\bs{\Omega}} \hat{\mathbf{A}}_{k}$ \\
		\hline
		\multirow{3}{*}{(1, 2)} & \multirow{3}{*}{18--41} & 
		$\hat{\bs{\Omega}} \hat{\mathbf{A}}_{p} \hat{\mathbf{S}}$, $\hat{\bs{\Omega}} \hat{\mathbf{A}}_{p} \hat{\mathbf{S}}^2$, $\hat{\bs{\Omega}}^2 \hat{\mathbf{A}}_{p} \hat{\mathbf{S}}$*, $\hat{\bs{\Omega}}^2 \hat{\mathbf{A}}_{p} \hat{\mathbf{S}}^2$*, $\hat{\bs{\Omega}}^2 \hat{\mathbf{S}} \hat{\mathbf{A}}_{p} \hat{\mathbf{S}}^2$*;\\
		
		&&$\hat{\bs{\Omega}} \hat{\mathbf{A}}_{k} \hat{\mathbf{S}}$,  $\hat{\bs{\Omega}} \hat{\mathbf{A}}_{k} \hat{\mathbf{S}}^2$,
		$\hat{\bs{\Omega}}^2 \hat{\mathbf{A}}_{k} \hat{\mathbf{S}}$*,  $\hat{\bs{\Omega}}^2 \hat{\mathbf{A}}_{k} \hat{\mathbf{S}}^2$*, $\hat{\bs{\Omega}}^2 \hat{\mathbf{S}} \hat{\mathbf{A}}_{k} \hat{\mathbf{S}}^2$*;\\
		
		&&$\hat{\mathbf{A}}_{p} \hat{\mathbf{A}}_{k} \hat{\mathbf{S}}$, $\hat{\mathbf{A}}_{p} \hat{\mathbf{A}}_{k} \hat{\mathbf{S}}^2$,
		$\hat{\mathbf{A}}_{p}^2 \hat{\mathbf{A}}_{k} \hat{\mathbf{S}}$*, $\hat{\mathbf{A}}_{p}^2 \hat{\mathbf{A}}_{k} \hat{\mathbf{S}}^2$*, $\hat{\mathbf{A}}_{p}^2 \hat{\mathbf{S}} \hat{\mathbf{A}}_{k} \hat{\mathbf{S}}^2$*;\\
		\hline
		(0, 3) & 42 & $\hat{\bs{\Omega}} \hat{\mathbf{A}}_{p} \hat{\mathbf{A}}_{k}$ \\
		\hline
		(1, 3) & 43--47 &   $\hat{\bs{\Omega}} \hat{\mathbf{A}}_{p} \hat{\mathbf{A}}_{k} \hat{\mathbf{S}}$,  $\hat{\bs{\Omega}} \hat{\mathbf{A}}_{k} \hat{\mathbf{A}}_{p} \hat{\mathbf{S}}$,  $\hat{\bs{\Omega}} \hat{\mathbf{A}}_{p} \hat{\mathbf{A}}_{k} \hat{\mathbf{S}}^2$,
		$\hat{\bs{\Omega}} \hat{\mathbf{A}}_{k} \hat{\mathbf{A}}_{p} \hat{\mathbf{S}}^2$,  $\hat{\bs{\Omega}} \hat{\mathbf{A}}_{p} \hat{\mathbf{S}} A_3 \mathbf{S}^2$ \\
		\hline						 								
	\end{tabular}
	\flushleft
	{\small
		Note: (a) The invariance basis is the trace of each tensor listed below.}	
\end{table}
Table~\ref{tab:basis} shows the minimal integrity bases for rotational invariance with given raw input 
set $\hat{\mathcal{Q}}$~\cite{spencer1962isotropic}. Note that the 
vectors $\nabla p$ and $\nabla k$ should be first mapped to antisymmetric tensors
 as follows,
 \begin{subequations}
 	\label{eq:vector2anti}
 	\begin{align}
 	\hat{\mathbf{A}}_p & =  -\mathbf{I} \times \nabla \hat{p}\\
 	\hat{\mathbf{A}}_k & =  -\mathbf{I} \times \nabla \hat{k}
 	\end{align}  	 
 \end{subequations}
 where $\mathbf{I}$ is the second order identity tensor, and $\times$ denotes tensor cross
 product. Note that the asterisk ($*$) on a term means to include all terms formed by cyclic
 permutation of anti-symmetric tensor labels (e.g., $\hat{\bs{\Omega}}^2 \hat{\mathbf{A}}_{p} \hat{\mathbf{S}}$* is short for $\hat{\bs{\Omega}}^2 \hat{\mathbf{A}}_{p} \hat{\mathbf{S}}$ and $ \hat{\mathbf{A}}_{p}^2 \hat{\bs{\Omega}} \hat{\mathbf{S}}$). As a result, a set of 47 invariant features is constructed to represent the information 
 of the mean flow feature set $\mathcal{Q}$. This construction procedure of input features 
 ensures completeness of rotational invariants with respect to the given set of raw tensor and vector inputs. 
 To further enrich the input features, this basis of 47 features from vector and tensor local 
 flow variables is supplemented by an additional ten features from Wang et al.~\cite{Wang2016}, 
 which also utilizes scalar RANS flow variables. For example, wall distance based Reynolds number 
 $Re_d$ is an important indicator to distinguish the boundary layer from shear flows. Although 
 some features in~\cite{Wang2016} may be redundant, since they are invariant functions of the 
 constructed invariant basis, the performance of random forest regressor is robust in the presence 
 of the redundant inputs. Finally, an input feature space of 57 invariants (collectively denoted as 
 $\mathbf{q}$) is constructed for machine learning.  
 
 \subsection{Representation of Reynolds Stress Discrepancy}
 As discussed in Section~\ref{sec:intro}, it is preferable to choose physical variables as the 
 responses (target results) of machine learning model. Since a major source of model-form errors in RANS 
 simulation comes from the modeled Reynolds stresses, a natural choice would be to directly 
 learn the Reynolds stresses from the data. However, this choice would totally abandon RANS 
 model and solely rely on data instead. Although potential model-form errors may exist, 
 the RANS model predictions are still valuable in most circumstances, and machine learning 
 should play the role as a complement instead of a replacement of RANS modeling. Therefore, 
 the discrepancies of RANS-modeled Reynolds stresses are suitable candidate for the responses. 
 Nevertheless, the discrepancies cannot be simply represented by the difference of each tensor 
 component, since it is frame dependent and difficult to impose physical constraints. Following 
 the work of Iaccarino and co-worker~\cite{emory2013modeling}, the discrepancies are formulated 
 in the six physically interpretable dimensions (i.e., magnitude, shape, and orientation) of Reynolds 
 stress tensor based on eigen-decomposition,
 \begin{equation}
 \label{eq:tau-decomp}
 \boldsymbol{\tau} = 2 k \left( \frac{1}{3} \mathbf{I} +  \mathbf{a} \right)
 = 2 k \left( \frac{1}{3} \mathbf{I} + \mathbf{V} \Lambda \mathbf{V}^T \right)
 \end{equation} 	
 where $k$ is the turbulent kinetic energy, which indicates the magnitude of $\bstau$; $\mathbf{I}$
 is the second order identity tensor; $\mathbf{a}$ is the deviatoric part of $\bstau$; 
 $\mathbf{V} = [\mathbf{v}_1, \mathbf{v}_2, \mathbf{v}_3]$ and $\Lambda = \textrm{diag}[\lambda_1, \lambda_2, \lambda_3]$ with $\lambda_1+\lambda_2+\lambda_3=0$
 are the orthonormal eigenvectors and eigenvalues of $\mathbf{a}$, respectively, indicating its shape
 and orientation. 

To impose the realizability constrain of Reynolds stress, the eigenvalues $\lambda_1$, $\lambda_2$,
and $\lambda_3$ are transformed to barycentric coordinates $C_1$, $C_2$, and $C_3$ as 
follows~\cite{banerjee2007presentation}:
 \begin{subequations}
 	\label{eq:lambda2c}
 	\begin{align}
 		C_1 & = \lambda_1 - \lambda_2 \\
 		C_2 & = 2(\lambda_2 - \lambda_3) \\
 		C_3 & = 3\lambda_3 + 1
 	\end{align}  
 \end{subequations}
 with $C_1 + C_2 + C_3 = 1$. The barycentric coordinates can be plotted as a triangle and 
 $C_1$, $C_2$, and $C_3$ indicate the portion of areas of the three sub-triangles in a 
 Cartesian coordinate $\bs\xi \equiv (\xi, \eta)$. Any point within the triangle is a convex
 combination of three vertices, i.e.,
 \begin{equation}
 \boldsymbol{\xi} = 	\boldsymbol{\xi}_{1c}C_1 + \boldsymbol{\xi}_{2c}C_2 +
 \boldsymbol{\xi}_{3c}C_3 
 \end{equation}
 where $\boldsymbol{\xi}_{1c}$, $\boldsymbol{\xi}_{2c}$, and $\boldsymbol{\xi}_{3c}$ denote
 coordinates of the three vertices of the triangle. After the above mapping, the coordinate 
 $\boldsymbol{\xi} \equiv (\xi, \eta)$ uniquely identifies the shape of the anisotropy tensor.
  Similar to the Lumley triangle~\cite{pope2001turbulent}, the Reynolds stresses falling in the 
  interior of the barycentric triangle are realizable. 
 
Representation of the discrepancy in the orientation (eigenvectors) of Reynolds stress tensor is
more challenging than that for eigenvalues. Moreover, there are no explicit physical constraints
on eigenvector systems. Less attention has been given to the quantification and reduction
of discrepancies in eigenvector systems. Wang et al.~\cite{mfu5} proposed a method to perturb
eigenvectors by using the Euler angle. Wang et al.~\cite{Wang2016} predicted the discrepancies
in eigenvectors parameterized by the Euler angle based on DNS data. In this work, the Euler 
angle is also employed to represent the discrepancies in eigenvectors of RANS-modeled Reynolds
stresses. The Euler angle system used follows the $z$--$x'$--$z''$ convention in rigid body 
dynamics~\cite{goldstein80euler}. That is, if a local coordinate system $x$--$y$--$z$ spanned by 
the three eigenvectors was initially aligned with the global coordinate system ($X$--$Y$--$Z$), 
the current configuration could be obtained by the following three consecutive intrinsic rotations 
about the axes of the local coordinate system: (1) a rotation about the $z$ axis by angle $\varphi_1$, 
(2) a rotation about the $x$ axis by $\varphi_2$, and (3) another rotation about its $z$ axis by 
$\varphi_3$.  The local coordinate axes usually change orientations after each rotation.  

Finally, the Reynolds stress tensor is projected to six physically meaningful parameters 
representing its magnitude ($k$), shape ($\xi$, $\eta$), and orientation ($\varphi_1$, 
$\varphi_2$, $\varphi_3$). The discrepancies ($\Delta \log k$, $\Delta \xi$, $\Delta \eta$, 
$\Delta \varphi_1$, $\Delta \varphi_2$, $\Delta \varphi_3$, collectively denoted as $\Delta{\bs\tau}$)
of Reynolds stress can be represented in these six projections. Note that the TKE 
discrepancy $\Delta\log k$ is the logarithm of the ratio of the DNS-predicted TKE ($k^{dns}$) to the 
RANS-predicted TKE ($k^{rans}$), i.e.,
\begin{equation}
\Delta\log k = \log{\frac{k^{dns}}{k^{rans}}}.
\end{equation}
Therefore, these discrepancies are dimensionless quantities. Moreover, they are also demonstrated to 
have similar characteristics among closely related flows~\cite{mfu3,Wang2016}, and thus are chosen as 
the learning targets. 
 
 \subsection{Construction of Machine Learning Model Based on Random Forest}
 After identifying the input feature set $\mathbf{q}$ and output responses $\Delta\bs\tau$, 
 a machine learning algorithm needs to be chosen to build the functional relation between
 input features and responses. There are various choices of supervised learning algorithms,
 e.g.,  K-nearest neighbors~\cite{altman1992introduction}, linear regression~\cite{james2013introduction}, 
 Gaussian processes~\cite{rasmussen2006gaussian}, tree-based methods (e.g., decision trees, random 
 forest, and bagging)~\cite{breiman2001random}, and neural networks~\cite{anderson1995introduction}. 
 As discussed in~\cite{Wang2016}, a major consideration is the capability to deal with the high 
 dimensionality of the feature space. This consideration becomes more important in the current work since the 
 input space is expanded to one with 57 features, the dimension of which is much higher than 
 that in~\cite{Wang2016}. Therefore, the random forest regressor~\cite{breiman2001random}, 
 known to be suitable for high dimensional regression problems, is employed to build the regression function 
 of Reynolds stress discrepancies with respect to the mean flow features. Random forest 
 regression is an ensemble learning technique that aggregates predictions from a number of decision trees. 
 The decision tree model stratifies the input space by a series of decision rules, and those rules can 
 be learned from the training data. By stratification of input space in a tree-like manner, the 
 decision tree is able to handle high-dimensional problems and is computationally efficient, as well. 
 However, one major disadvantage of the single decision tree is that it tends to overfit the data, which 
 often leads to poor predictions. This disadvantage can be overcome and prediction performance can be 
 significantly improved by aggregating a large number of trees, which is the essence of the random forest 
 algorithm. In the random forest model, the ensemble of decision trees is built with bootstrap aggregation 
 samples (i.e., sampling with replacement) drawn from the training data~\cite{friedman2001elements}.  
 Moreover, only a subset of randomly selected features is used when determining the splits of each 
 tree. This reduces the correlation among trees in the ensemble and thus decreases the bias of 
 the ensemble prediction.
                                                                                                                              
\section{Numerical Results}
\label{sec:result}
\subsection{Case Setup}
\begin{figure}[htbp]
	\centering
	\includegraphics[width=0.8\textwidth]{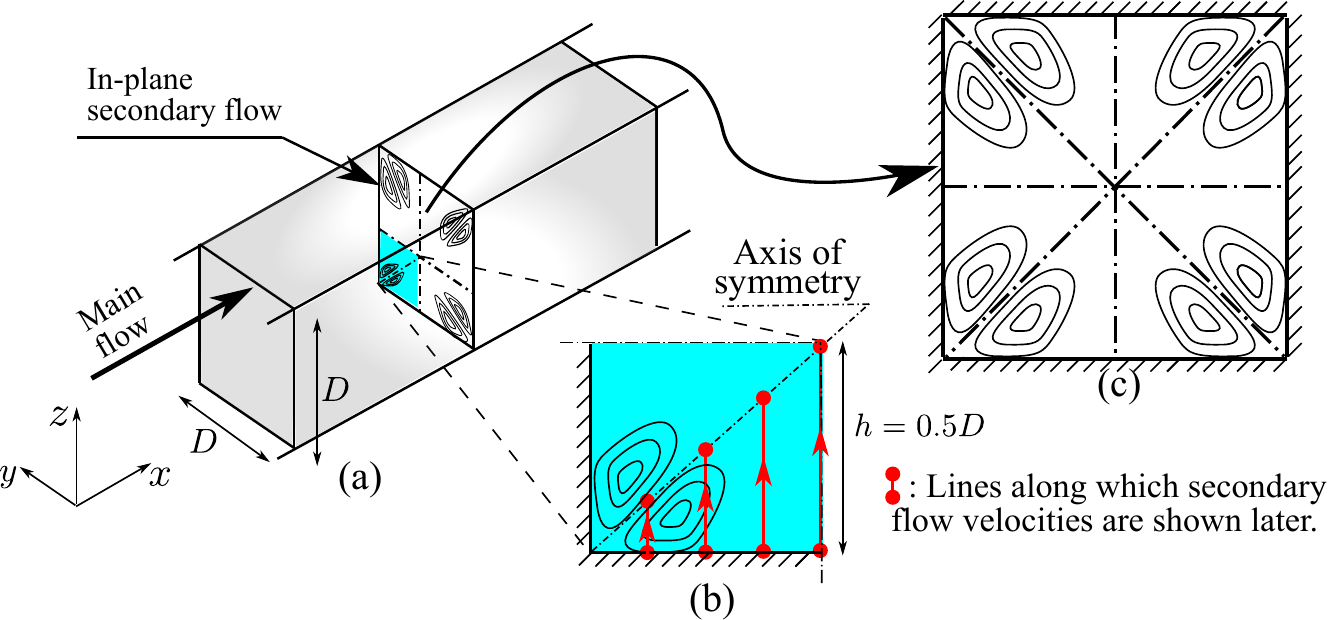}
	\caption{Domain shape for the flow in a square duct. The $x$ coordinate represents the streamwise
		direction. Secondary flows induced by Reynolds stress imbalance exist in the $y$--$z$
		plane. Panel (b) shows that the computational domain covers a quarter of the cross-section of the
		physical domain. This is due to the symmetry of the mean flow in both $y$ and $z$ directions as
		shown in panel (c).}
	\label{fig:domain_duct}
\end{figure}

The fully developed turbulent flow in a square duct is considered to demonstrate the proposed framework. 
Although this flow has a simple geometry, it features the in-plane secondary flow induced by Reynolds 
stresses. All RANS turbulence models under linear eddy viscosity hypothesis fail to predict the secondary 
mean motion, and even the Reynolds-stress transport model (RSTM) cannot predict it well~\cite{billard2011}. 
The errors stem from the modeled Reynolds stresses. Therefore, we aim to improve the RANS-predicted 
Reynolds stresses by learning from the data of similar flows. The geometry of the duct flow is presented 
in Fig.~\ref{fig:domain_duct}. The Reynolds number $Re$ is based on the edge length $D$ of the square 
and bulk velocity. All lengths presented below are normalized by the height $h$ of the computational 
domain, which is half of the edge length. 
 
The Reynolds stress discrepancy function is trained on the data from flows at Reynolds numbers 
$Re = 2200, 2600, 2900$ to predict the flow at a higher Reynolds number $Re = 3500$. All flows 
have the same geometry. The data of training flows are obtained from direct numerical simulations 
(DNS)~\cite{pinelli2010reynolds}. Note that the DNS data of the flow to be predicted ($Re = 3500$) 
are reserved for comparison and are not used for training. The mean flow patterns among these 
flows are similar. In the cross-plane of the duct, there is a counter-rotating pair located at each of 
the four corners (Fig.~\ref{fig:domain_duct}(c)). However, the recirculation bubble moves closer to 
the wall and its size decreases as the Reynolds number increases. 
 
Baseline RANS simulation is conducted for each flow to obtain mean flow features and training data 
of the Reynolds stress discrepancies. Since linear eddy viscosity models are not able to predict the 
mean flow features of the secondary motions, the Launder-Gibson RSTM is adopted to perform 
the baseline simulation. As indicated in Fig.~\ref{fig:domain_duct}, only one quadrant of the physical 
domain is simulated due to the symmetry of the mean flow with respect to the centerlines along 
$y$- and $z$-axes. No-slip boundary conditions are applied at the walls, and symmetry boundary 
conditions are applied on the symmetry planes. The DNS Reynolds stresses are interpolated onto the 
mesh of the RANS simulation to calculate the discrepancy. The RANS simulations are performed in an 
open-source CFD platform, OpenFOAM, using a built-in incompressible flow 
solver~\texttt{simpleFoam}~\cite{weller1998tensorial}. Mesh convergence studies have been performed.

The random forest regressor is constructed with decision trees, each of which is built to its 
maximum depth by successive splitting of nodes until each leaf is left with one training data point. 
To control the prediction performance of a random forest model, there are two important free 
parameters, i.e., number of trees and number of randomly selected features on which each split on 
each tree is determined. Generally, a higher number of trees leads to a better performance. Based on our testing, 
an ensemble of $200$ trees is large enough to have a robust prediction. The number of features randomly 
selected is commonly smaller than the total number of input features. The reason for embedding 
the randomness is to enhance the diversity of the trees. Therefore, the random forest prediction can be 
more robust and unlikely to over-fit the data. The size of the randomly selected subset of features 
is commonly chosen as the square root of the total number of input features~\cite{svetnik2003random}.
In the current test case, the prediction results were shown to be insensitive to the number of features over
which each split was determined.

\subsection{Results and Interpretation}
\subsubsection{Verification of DNS Data}
The aim of the PIML framework is to reduce the discrepancies in the RANS-modeled Reynolds 
stresses. With the improved Reynolds stresses, one should be able to obtain an accurate prediction 
of the velocity field. However, the outcome of velocity propagation depends on the quality of 
training data, i.e., DNS Reynolds stresses. Although the Reynolds stresses from DNS simulations 
are assumed to be more accurate than the RANS predictions, it is not guaranteed that they can 
be propagated to a better mean velocity field due to potential statistical convergence errors. 
Thompson et al.~\cite{thompson2016methodology} recently demonstrated that, even for channel 
flows, the Reynolds stresses of different DNS databases in literature lead to significant discrepancies 
in the propagated velocity fields. To better evaluate the performance of machine 
learning predictions of propagated velocities, it is useful to check the velocity field obtained by 
directly propagating the DNS Reynolds stresses via RANS equations. 
\begin{figure}[htbp]
	\centering
	\includegraphics[width=0.35\textwidth]{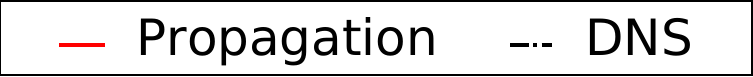}\\ 
	\subfloat[$U_y$]{\includegraphics[width=0.45\textwidth]{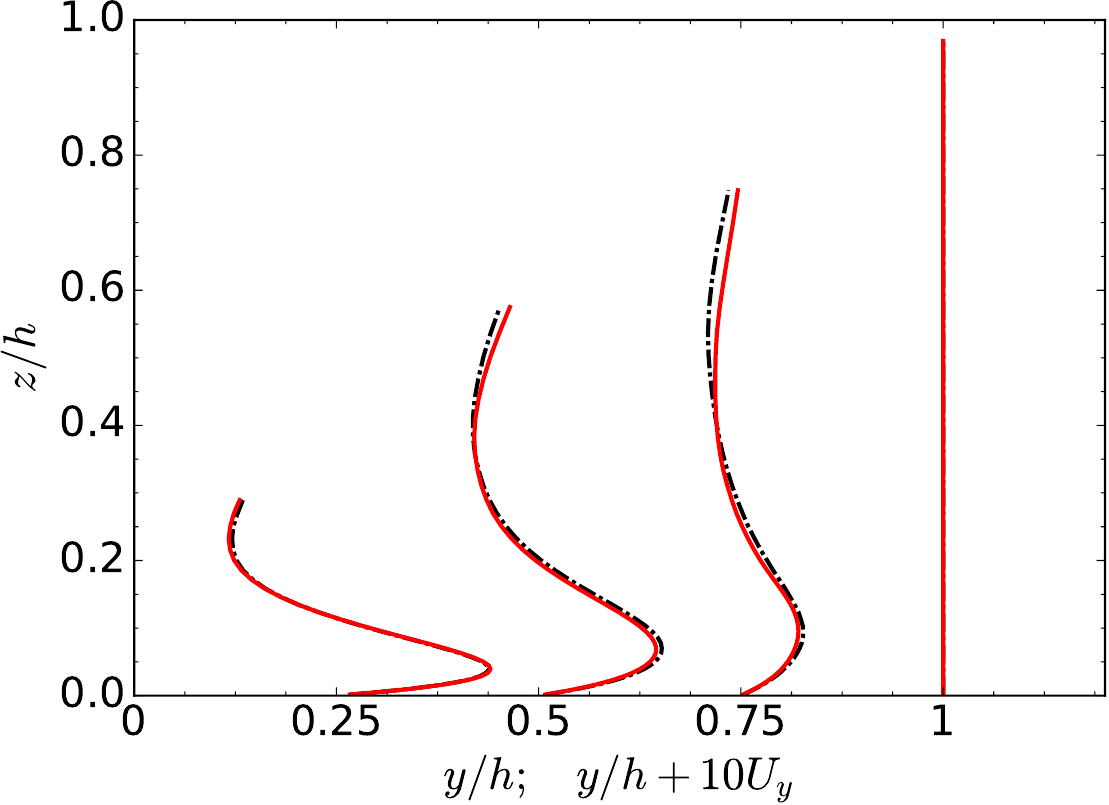}} 
	\subfloat[$U_z$]{\includegraphics[width=0.45\textwidth]{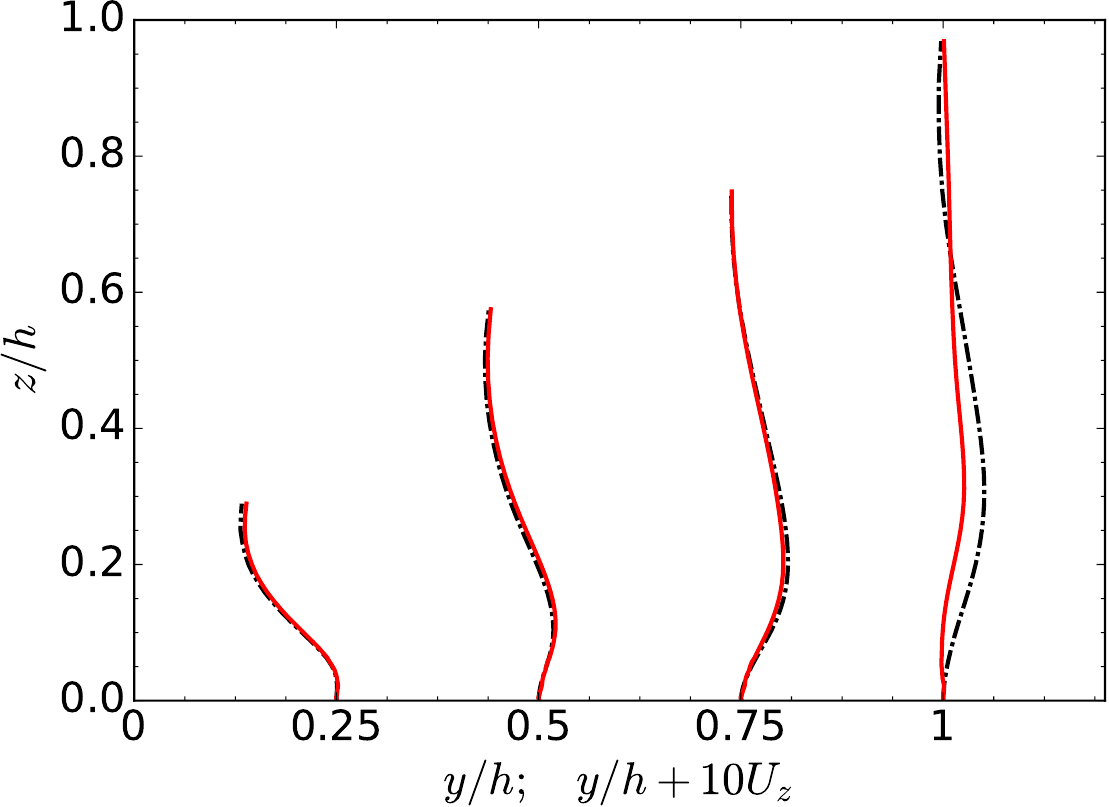}}
	\caption{In-plane velocity profiles (a) $U_y$ and (b) $U_z$ obtained by 
		propagating the DNS Reynolds stresses $\bstau_{dns}$ via RANS 
		equations at Reynolds number $Re = 3500$. The DNS results are also plotted for comparison.}				
	\label{fig:U_DNS}
\end{figure}

The DNS data of Reynolds stresses at $Re = 3500$ are used to solve the RANS equations, and 
the corresponding mean velocity field is obtained. The propagated mean field of in-plane velocity 
is compared to that provided by the DNS space-time averaging, which is shown in Fig.~\ref{fig:U_DNS}. 
The in-plane velocity components $U_y$ and $U_z$ on the four cross-sections ($y/h = 0.25, 0.5, 0.75$ 
and $1.0$, as indicated in Fig.~\ref{fig:domain_duct}(b)) are presented, but only the profiles in the 
region below the diagonal are shown due to the diagonal symmetry. It can be seen that the propagated 
results agree well with the DNS profiles along all four cross sections for both $U_y$ and $U_z$. However, 
in the regions away from the corner (e.g., $y/h > 0.75$ or $z/h > 0.4$), the propagated velocity profiles 
slightly deviate from the DNS results. Especially for $U_z$, notable discrepancies can be observed in the 
profile at $y/h = 1$. These discrepancies might come from small errors in Reynolds stresses (e.g., error introduced in interpolations to RANS mesh), since the secondary flow is sensitive to the Reynolds stress 
components. As the magnitude of secondary velocity decreases away from the corner, the discrepancies 
are even more pronounced. Nevertheless, the overall data quality of Reynolds stresses are considered 
satisfactory to obtain an improved velocity field. 

\subsubsection{Learning and Prediction of Reynolds Stress}
The discrepancy functions in six physical projections are learned from the training flows, 
and predictions are made for the test flow. We first investigate the prediction performance 
of the shape of the Reynolds stress anisotropy tensor, which can be visualized in a barycentric 
triangle. 
\begin{figure}[htbp]
	\centering 
	\subfloat[$x/H = 0.25$]{\includegraphics[width=0.45\textwidth]{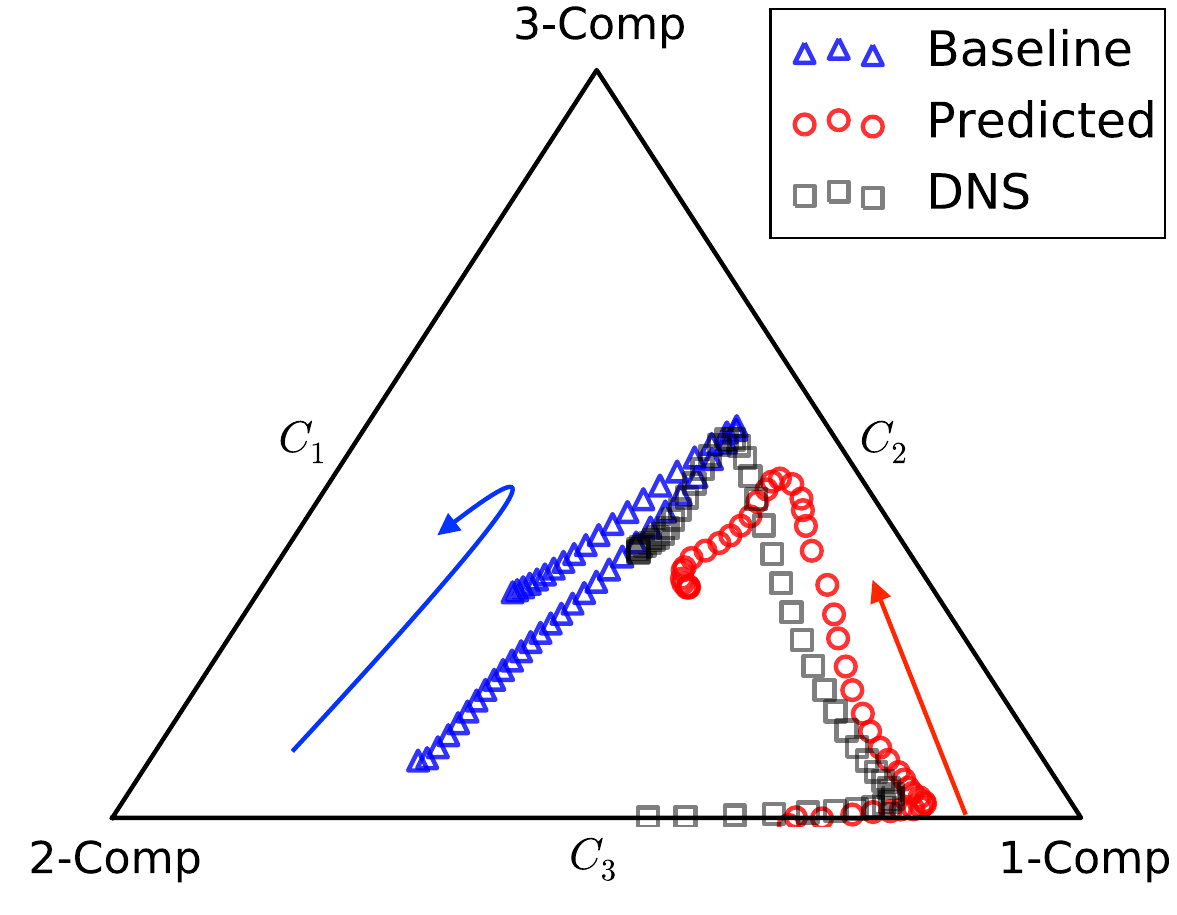}} 
	\subfloat[$x/H = 0.75$]{\includegraphics[width=0.45\textwidth]{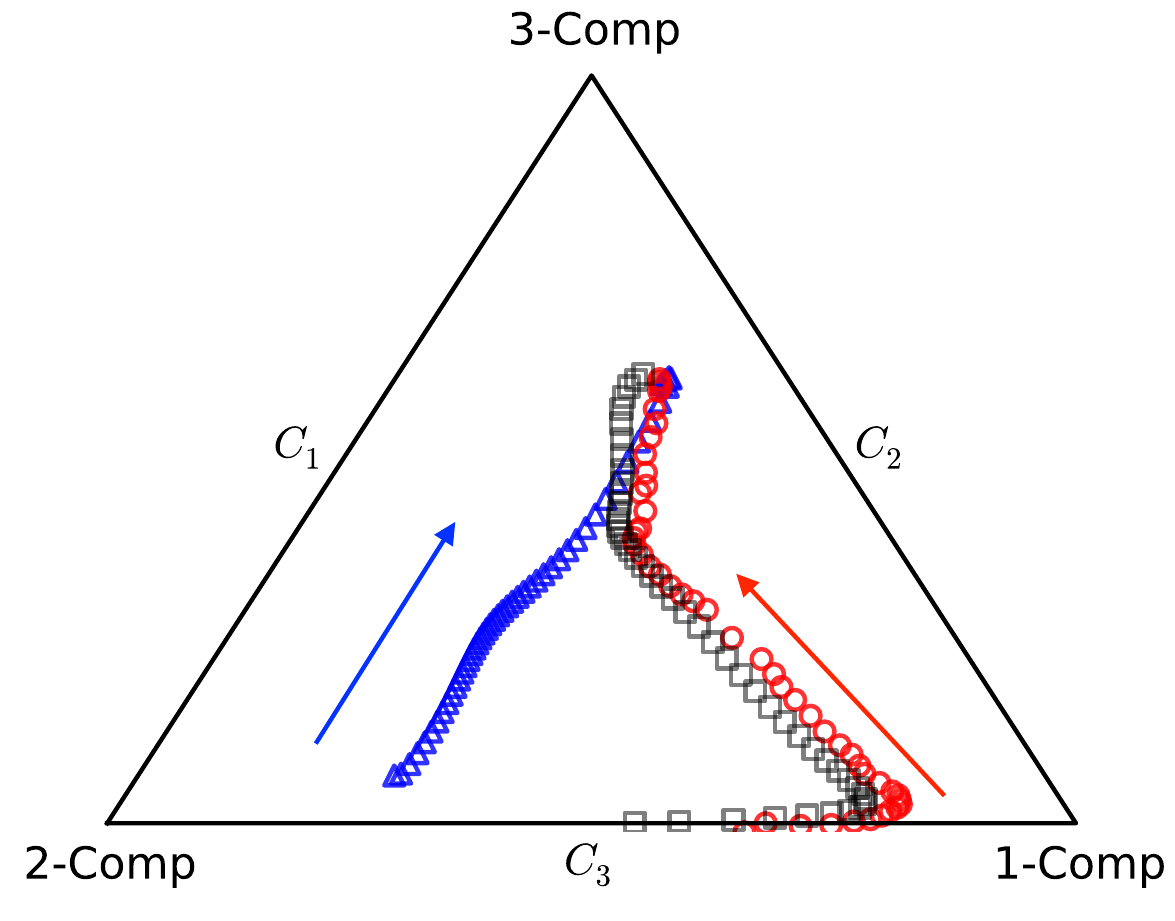}}
	\caption{Barycentric map of the predicted Reynolds stress anisotropy for the test flow 
		($Re = 3500$), learned from the training flows ($Re = 2200, 2600$, and $2900$)
		The prediction results on two streamwise locations at $x/H = 0.25$ and $0.75$ are compared with
		the corresponding baseline (RSTM) and DNS results in panels (a) and (b), 
		respectively.}
	\label{fig:bayRe}
\end{figure}
The prediction results on two typical cross-sections ($y/H = 0.25$ and $y/H = 0.75$) are plotted in 
the barycentric triangle in Figs.~\ref{fig:bayRe}a and~\ref{fig:bayRe}b, respectively. The Reynolds 
stresses on the cross-section at $y/H = 0.25$ from the wall to the outer layer start from two-component 
limit states (bottom edge) towards three-component anisotropic states (middle area of the triangle). 
For the cross-section at $y/H = 0.75$, the spatial variation of turbulence states is similar. The baseline 
RSTM results capture this trend to some extent, especially in the regions away from the wall. This does 
much better than the linear eddy viscosity models (results are not shown), predictions of which have an
opposite trend against the truth. Although the RSTM predictions away from the wall are satisfactory, 
discrepancies are still pronounced, especially in the near wall region. For example, 
It can be seen in Fig.~\ref{fig:bayRe}a that the DNS Reynolds stress anisotropy on the wall is
at the two component state, since the velocity fluctuations in the wall normal direction are suppressed 
by the blocking of the bottom wall. As away from the wall, it moves towards the one-component state 
and then towards the three-component anisotropic states. In contrast, RSTM-predicted anisotropy on 
the wall is closer to the two-component axi-symmetric state, while it approaches directly towards the 
generic turbulent states as away from the wall. By correcting the baseline RANS-modeled 
Reynolds stresses with the trained discrepancy function, the PIML-predicted anisotropy is significantly 
improved. In both Figs~\ref{fig:bayRe}a and~\ref{fig:bayRe}b, the PIML-predicted anisotropies (circles) 
show much better agreement with the DNS results (squares) than the RSTM prediction does, especially 
in the near wall regions. 

\begin{figure}[htbp]
	\centering
	\includegraphics[width=0.5\textwidth]{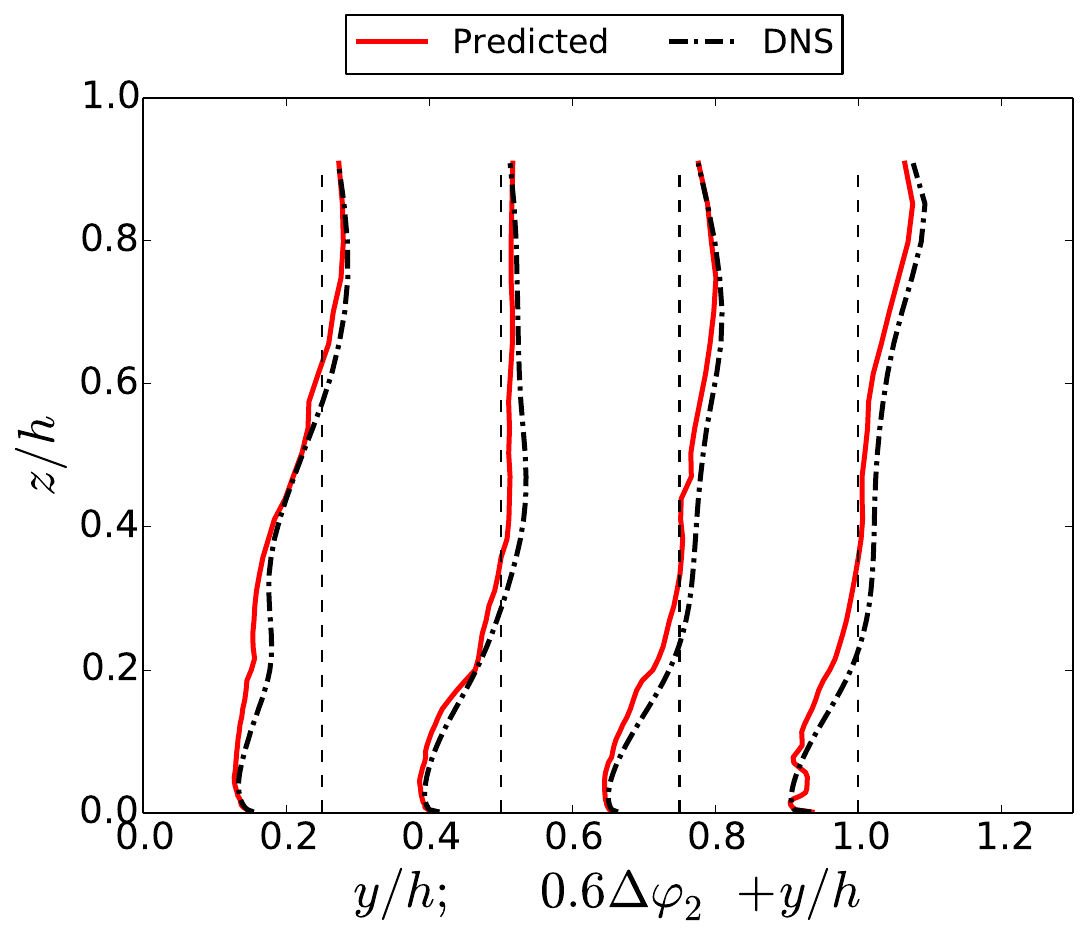}
	\caption{Rotation angle $\Delta \varphi_2$ of predicted Reynolds stress from the
		baseline for the test flow (Re = 3500). The profiles are shown at four streamwise 
		locations $x/H = 0.25, 0.5, 0.75,$ and $1$. Corresponding DNS and 
		baseline (reference lines) results are also plotted for comparison.}
	\label{fig:angle}
\end{figure}
The barycentric coordinates of anisotropy tensor are shown to be considerably improved in 
the machine learning prediction process. However, the improvement of eigenvalues alone may 
not necessarily lead to a better prediction of anisotropy tensor. When the eigenvectors of 
RANS-predicted anisotropy markedly deviate from the DNS results, the reconstructed Reynolds 
stress even with the DNS eigenvalues but the RANS eigenvectors may lead to even larger discrepancies in its tensor 
components. Therefore, corrections are required for both the shape and orientation of the 
anisotropy tensor. Figure~\ref{fig:angle} presents the discrepancy profiles of RANS-predicted 
Reynolds stresses in orientations, i.e., the rotation angles of DNS anisotropies from the 
baseline RSTM results. Note that only the angle discrepancies $\Delta\varphi_2$ are presented, and 
the results for $\Delta\varphi_1$ and $\Delta\varphi_3$ are omitted due to their qualitative 
similarities. The angle is presented in radian. Notable discrepancies between the eigenvector systems 
of the DNS and baseline anisotropy tensors can be observed. Especially, in the near wall 
regions ($y/H < 0.2$), the rotation angles $\delta\varphi_2$ are more than 0.2 rad ($10$ degrees). 
It can be seen that these angle discrepancies are well predicted by the trained regression function 
over the entire domain, and their spatial variations are also well captured. However, slight wiggles can 
be found in the predicted $\Delta\varphi_2$ on the lower part of the profile at $y/H = 1$. This 
non-smoothness originates from the pointwise estimation of the Random forest model in the feature space, 
which cannot guarantee spatial smoothness in the physical domain.

\begin{figure}[htbp]
	\centering
	\includegraphics[width=0.5\textwidth]{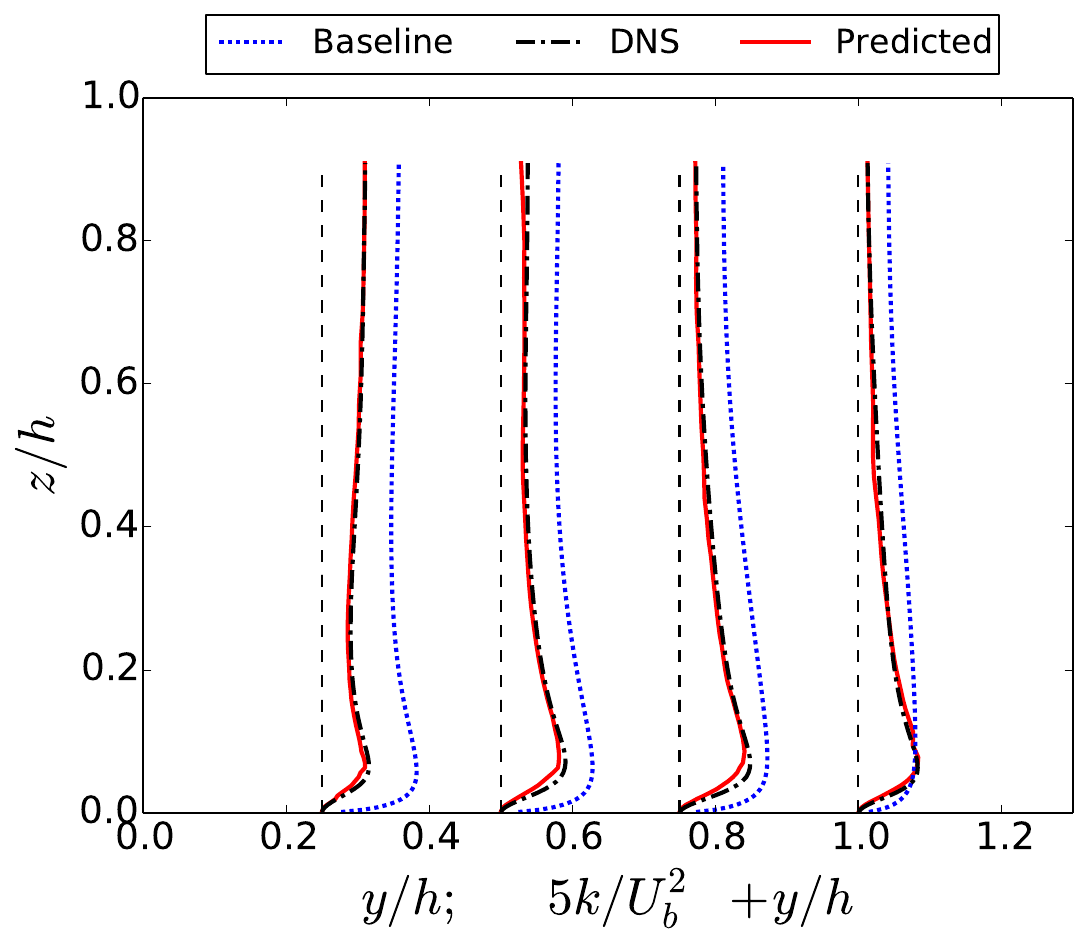}
	\caption{Turbulence kinetic energy of the test flow (Re = 3500), learned from the 
		training flows ($Re = 2200, 2600$, and $2900$). 
		The profiles are shown at four streamwise locations $x/H = 0.25, 0.5, 0.75,$ and $1$. 
		Corresponding DNS and baseline (RSTM) results are also plotted for comparison.}
	\label{fig:TKE}
\end{figure}
The TKE is also not well predicted by the baseline RANS model, which can be seen in Fig.~\ref{fig:TKE}. 
The RSTM tends to overestimate the magnitudes of Reynolds stresses, which are almost the twice those of 
the DNS results in most regions. The discrepancies of the RSTM modeled TKE exist in the entire flow 
domain and are especially large close to the corner. Similar to the anisotropy prediction, the TKE field 
corrected by the trained random forest model is significantly improved over the baseline results. 
Fig.~\ref{fig:TKE} shows that the TKE profiles of corrected Reynolds stresses are nearly identical to 
the DNS profiles.

\begin{figure}[htpb]
	\centering
	\includegraphics[width=0.5\textwidth]{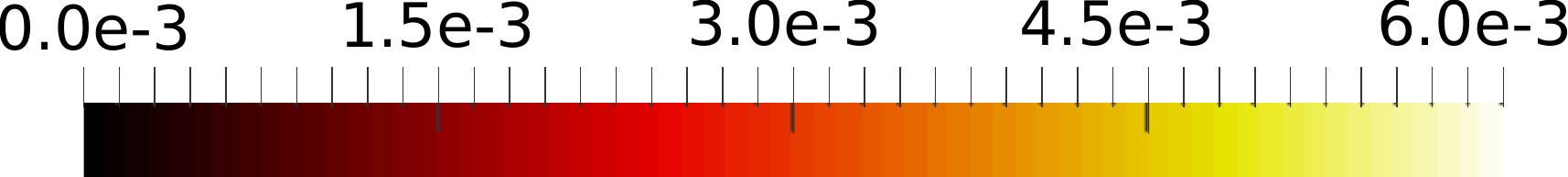}\\ 
	\subfloat[Baseline $\tau_{yy}$]{\includegraphics[width=0.3\textwidth]{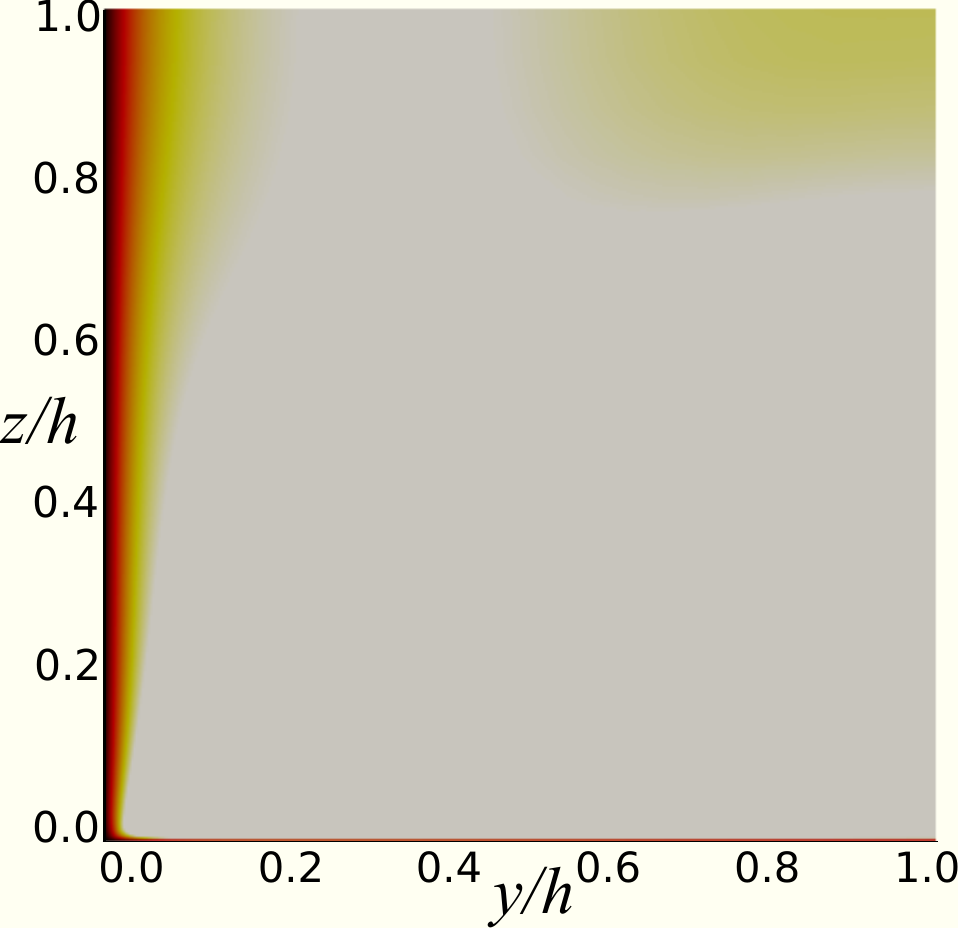}} \hfill
	\subfloat[DNS $\tau_{yy}$]{\includegraphics[width=0.3\textwidth]{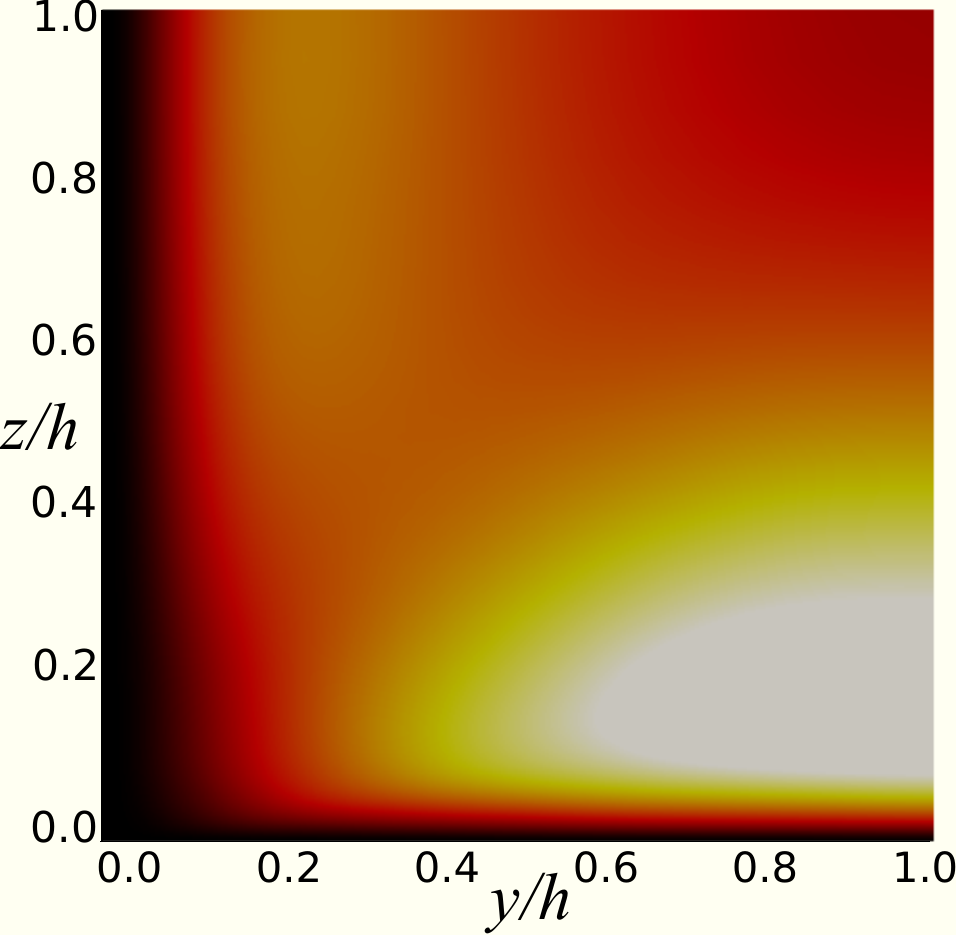}}\hfill
	\subfloat[Predicted $\tau_{yy}$]{\includegraphics[width=0.3\textwidth]{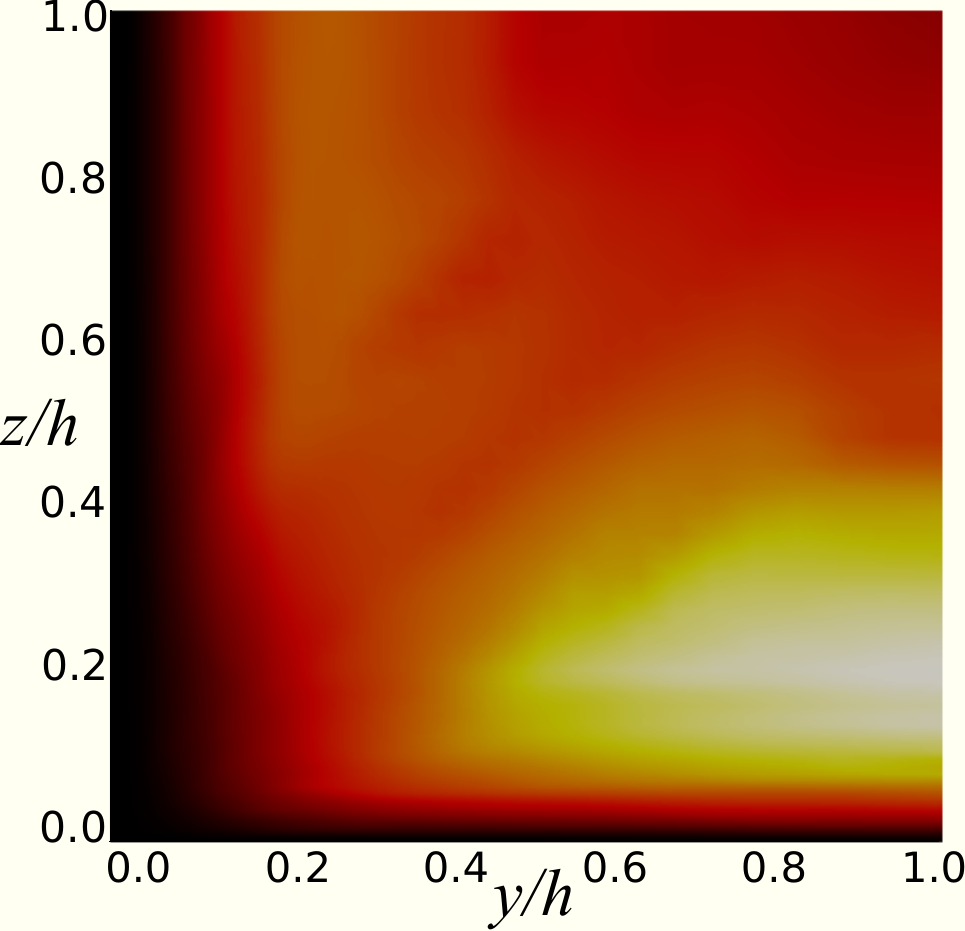}}\\
	\subfloat[Baseline $\tau_{zz}$]{\includegraphics[width=0.3\textwidth]{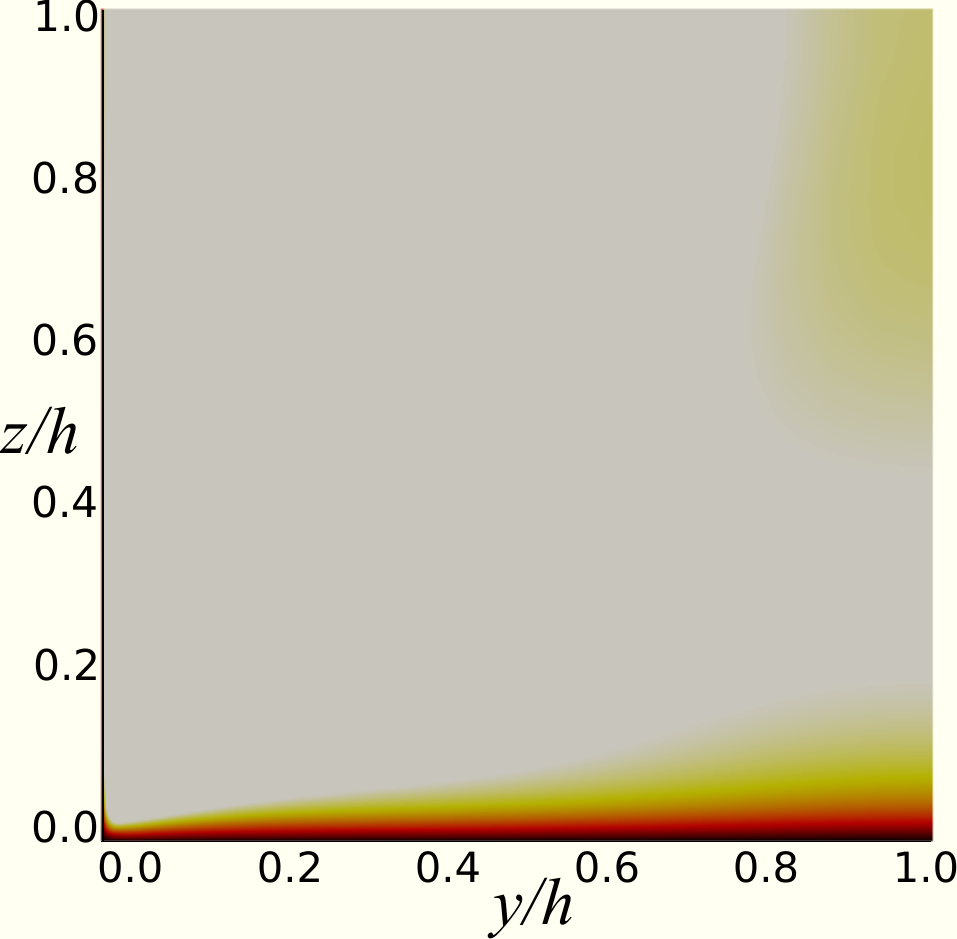}} \hfill
	\subfloat[DNS $\tau_{zz}$]{\includegraphics[width=0.3\textwidth]{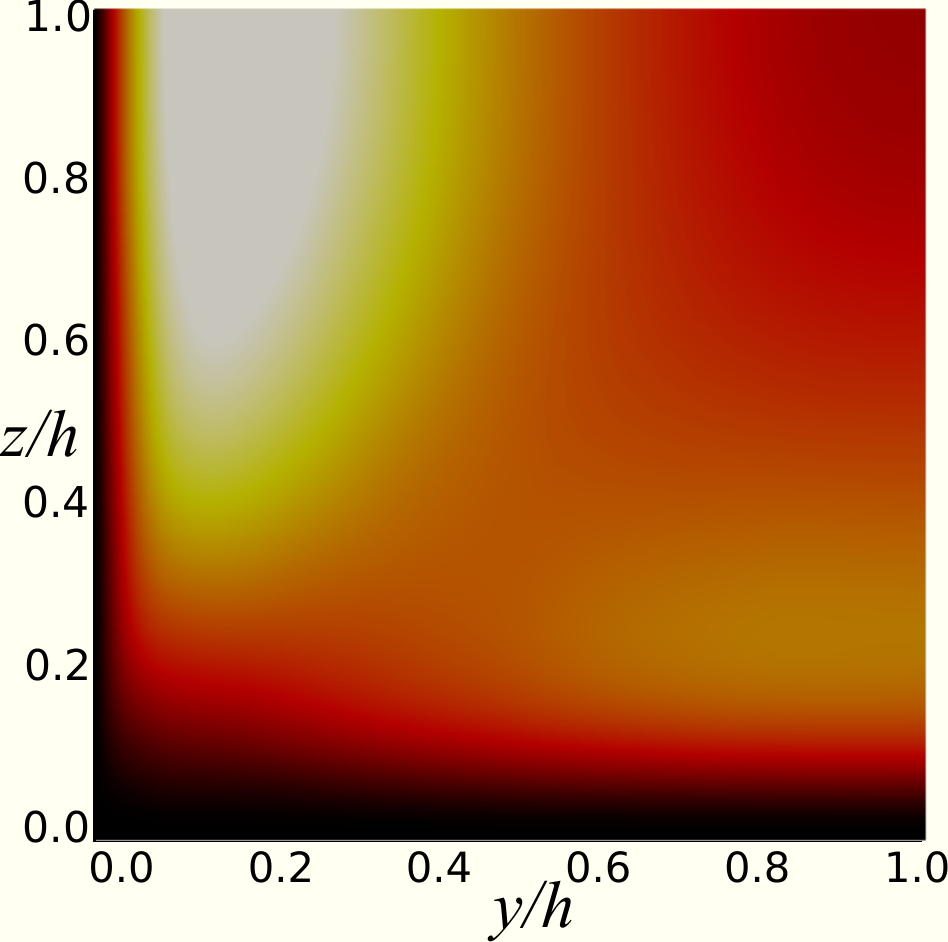}}\hfill
	\subfloat[Predicted $\tau_{zz}$]{\includegraphics[width=0.3\textwidth]{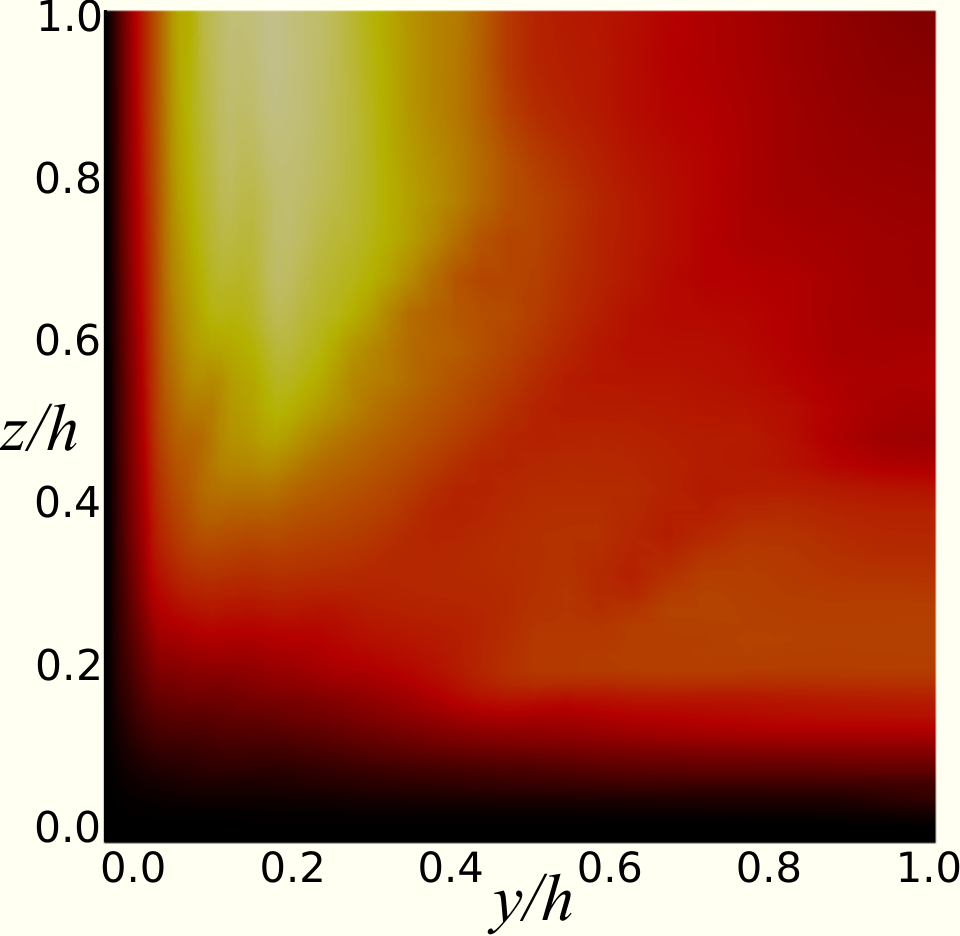}}		 
	\caption{Contour plots of normal components $\tau_{yy}$ and $\tau_{zz}$ for baseline (a, d), 
					DNS (b, e) and machine-learning-predicted (c, f) results.}
	\label{fig:tau_cont}
\end{figure}

\begin{figure}[htbp]
	\centering
	\includegraphics[width=0.4\textwidth]{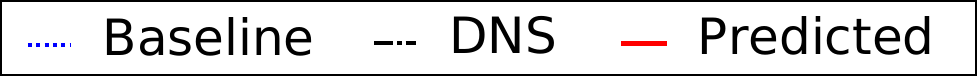}\\	
	\subfloat[normal stress imbalance]{\includegraphics[width=0.45\textwidth]{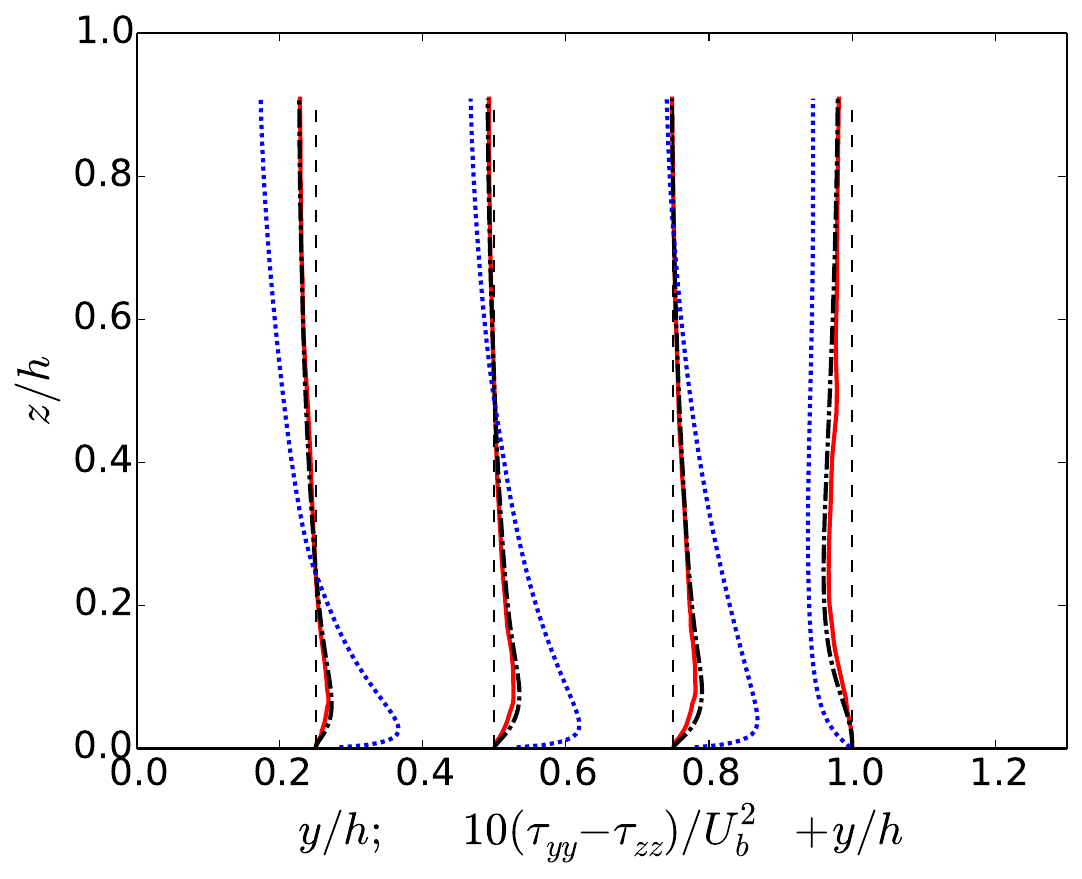}}
	\subfloat[shear component]{\includegraphics[width=0.45\textwidth]{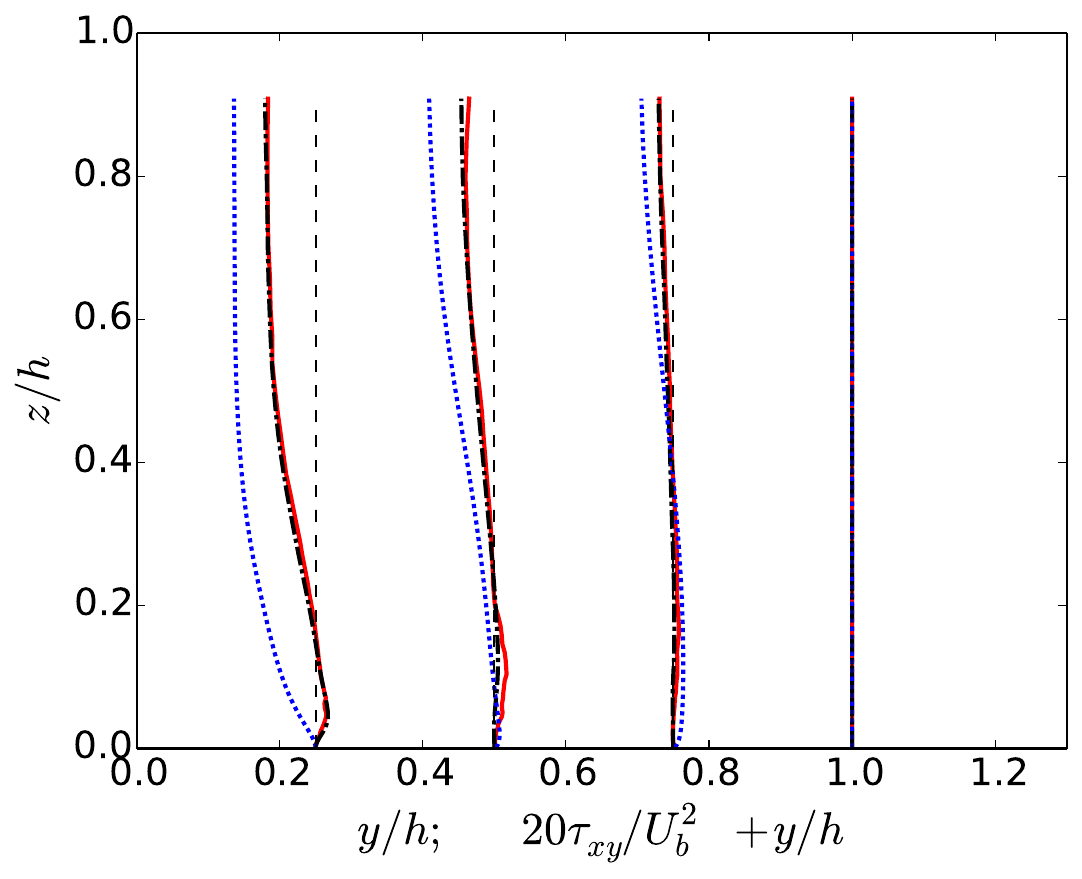}}
	\caption{Profiles of (a) normal stress imbalance $\tau_{yy} - \tau_{zz}$ and (b) shear component $\tau_{xy}$ 
		of corrected Reynolds stress with the discrepancy model trained on 57 features. 
		The profiles are shown at four streamwise locations $x/H = 0.25, 0.5, 0.75,$ and $1$. 
		Corresponding DNS and baseline (RSTM) results are also plotted for comparison.}
	\label{fig:tau}
\end{figure}

The results shown above demonstrate that all the physical projections of the RANS-predicted Reynolds 
stresses are significantly improved by the random forest discrepancy model. Therefore, it is expected 
that the tensor components should also be improved over the corresponding baselines. 
Figure~\ref{fig:tau_cont} shows the contour plot comparisons of the baseline, DNS, and PIML-predicted 
results on the turbulent normal stress components $\tau_{yy}$ and $\tau_{zz}$. These two normal 
components of Reynolds stress tensor are known to be important to the velocity propagation in the 
duct flow since the imbalance of them ($\tau_{yy} - \tau_{zz}$) is the main driving force of the 
secondary flow. Both $\tau_{yy}$ and $\tau_{zz}$ are markedly overestimated by the RSTM over the 
entire domain, which is due to its overestimation of the TKE (see Fig.~\ref{fig:TKE}). Moreover, the 
spatial variation patterns of RSTM predictions are significantly different from those of DNS results, 
especially in the near-corner region. As expected, the machine learning predictions are considerably 
improved over the RSTM baseline. Most of the magnitudes, features, and patterns of DNS results are 
captured well in the PIML predictions for both $\tau_{yy}$ and $\tau_{zz}$. In Fig.~\ref{fig:tau}a 
and~\ref{fig:tau}b, we also compare the profiles of normal stress imbalance $\tau_{yy} - \tau_{zz}$
and turbulent shear stress $\tau_{xy}$ on four cross sections to more clearly demonstrate the 
improvement in the machine-learning predictions. It can be seen that RSTM captures the spatial 
pattern of the normal stress imbalance, which is positive near the wall and becomes negative away 
from the wall. However, the magnitude $\|\tau_{yy} - \tau_{zz}\|$ of imbalance term is significantly 
overestimated. Moreover, the RSTM underestimates the turbulent shear stress $\tau_{xy}$. 
The discrepancies of RSTM-modeled shear component $\tau_{xy}$ are more notable on 
the cross-section at $y/H = 0.25$, which is close to the left lower corner. As expected, 
the PIML-corrected results shows pronounced improvements over the RSTM baselines. 
The PIML predictions nearly coincide with the DNS results for both $\tau_{yy} - \tau_{zz}$ 
and $\tau_{xy}$ profiles. This demonstrates that the discrepancies in all Reynolds stress tensor 
components that are relevant to the mean motion predictions are well predicted by the trained 
discrepancy functions.

\subsubsection{Propagation of Improved Reynolds Stress Prediction}
The improvement of Reynolds stresses enabled by the PIML framework, the success of which has been 
demonstrated above, is an important step toward data-driven, predictive turbulence modeling. 
However, the ultimate goal is to obtain more accurate quantities of interest (QoI) after propagating 
the corrected Reynolds stress through RANS equations. To investigate the performance on the 
improvement of the propagated mean velocity field, we substitute the Reynolds stress field of 
RANS momentum equations with the corrected one and propagate them by solving the equations.

\begin{figure}[htpb]
	\centering 
	\includegraphics[width=0.4\textwidth]{u_f57_legend}\\
	\subfloat[$U_y$]{\includegraphics[width=0.45\textwidth]{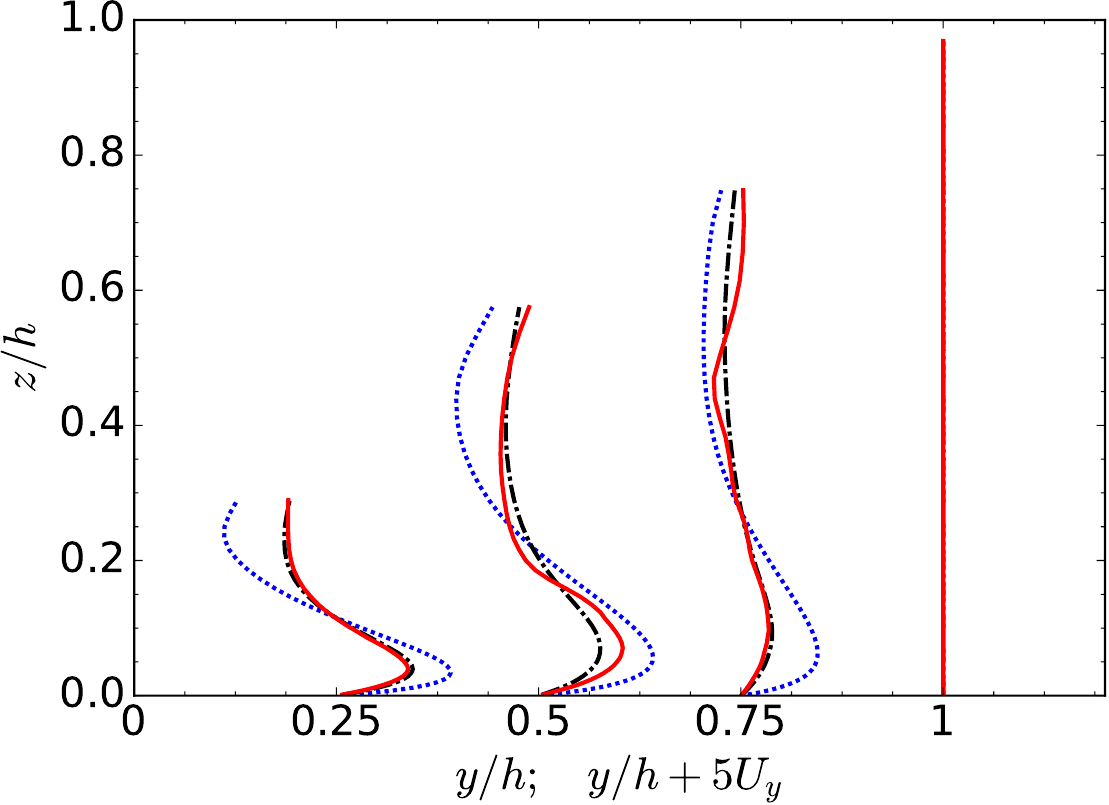}} 
	\subfloat[$U_z$]{\includegraphics[width=0.45\textwidth]{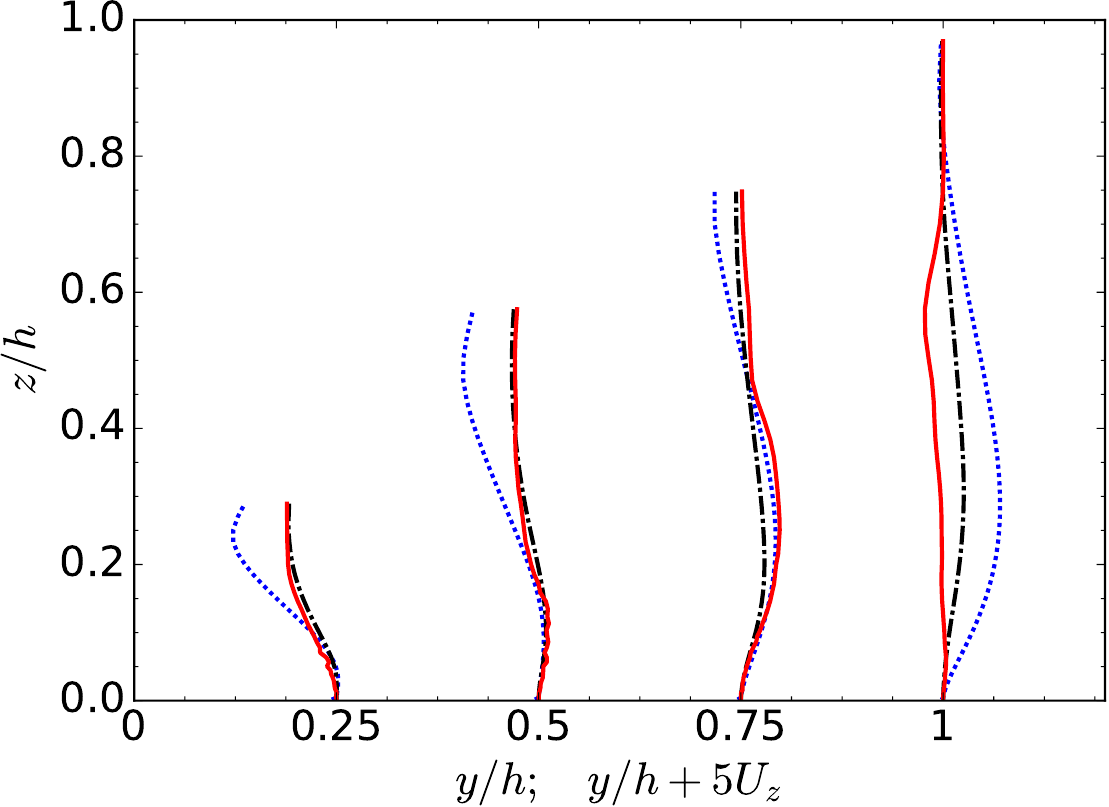}}
	\caption{In-plane velocity profiles (a) $U_y$ and (b) $U_z$ obtained by propagating the 
		PIML predicted Reynolds stresses via RANS equations at Reynolds number $Re = 3500$. 
		The baseline (RSTM) and DNS results are also plotted for comparison. 
		The discrepancy functions of Reynolds stresses are trained on 57 features. }
	\label{fig:U_f57}
\end{figure}
\begin{figure}[htpb]
	\centering
	\includegraphics[width=0.3\textwidth]{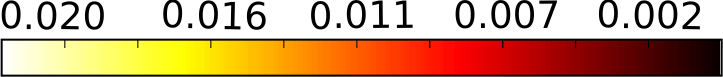}\\ 
	\subfloat[Baseline RSTM]{\includegraphics[width=0.33\textwidth]{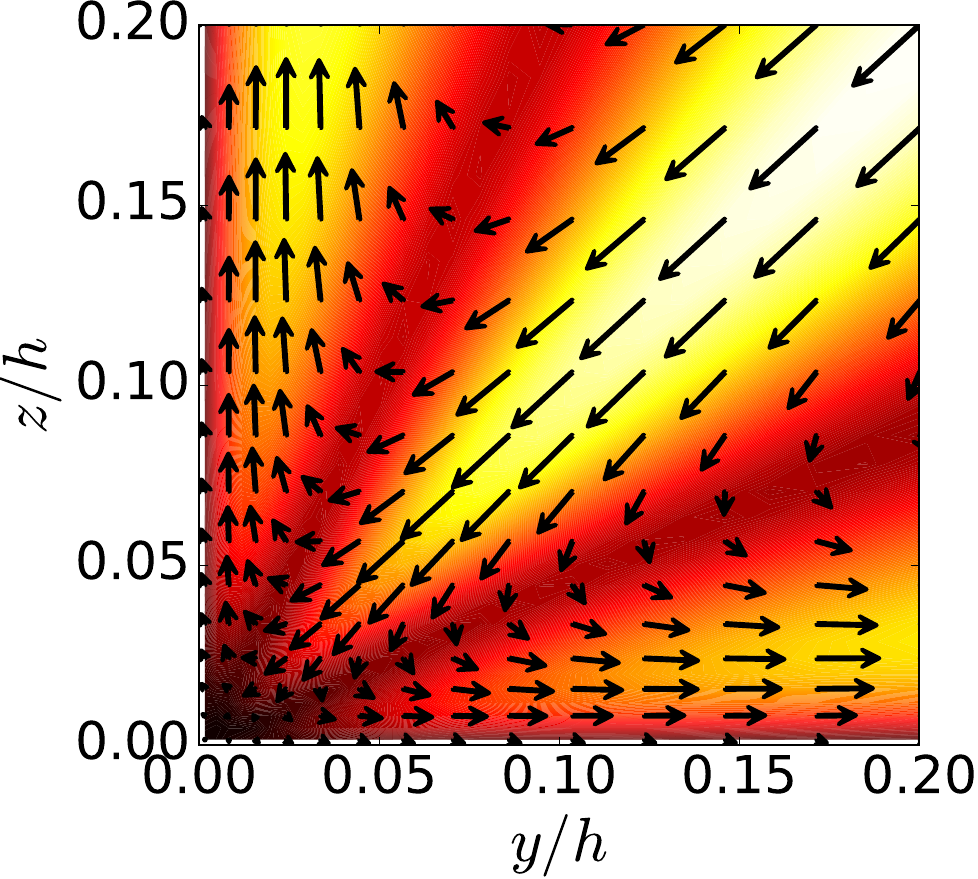}} 
	\subfloat[DNS]{\includegraphics[width=0.33\textwidth]{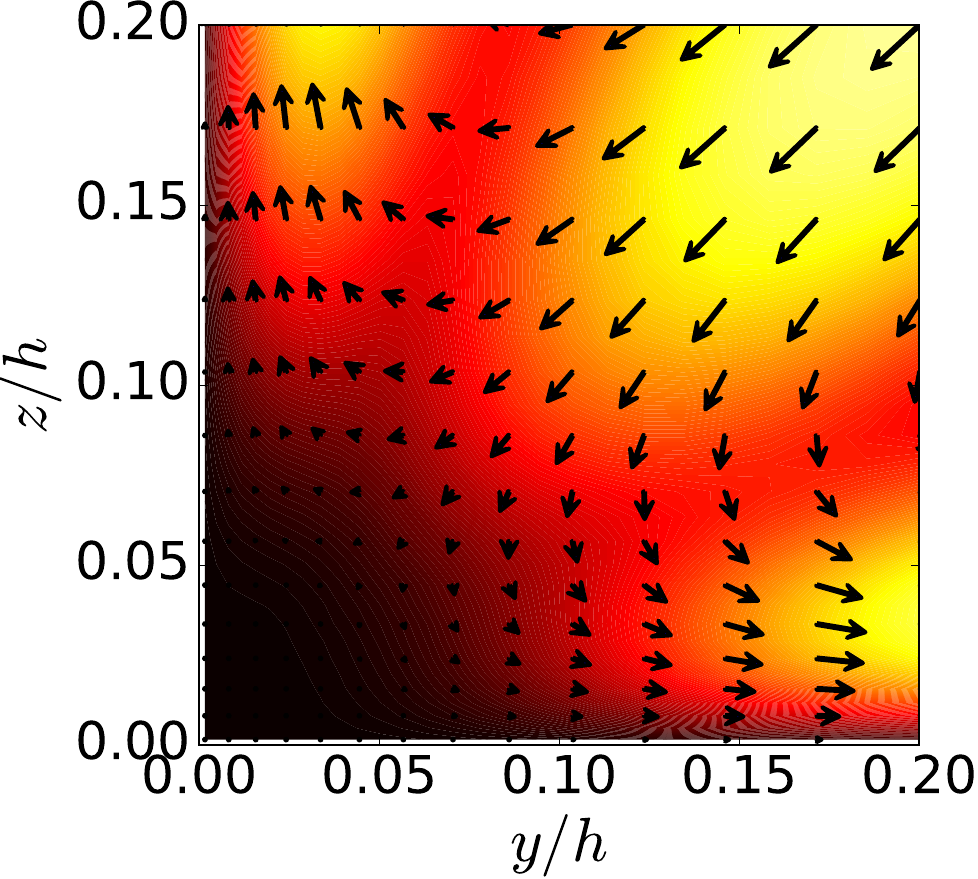}}
	\subfloat[Predicted]{\includegraphics[width=0.33\textwidth]{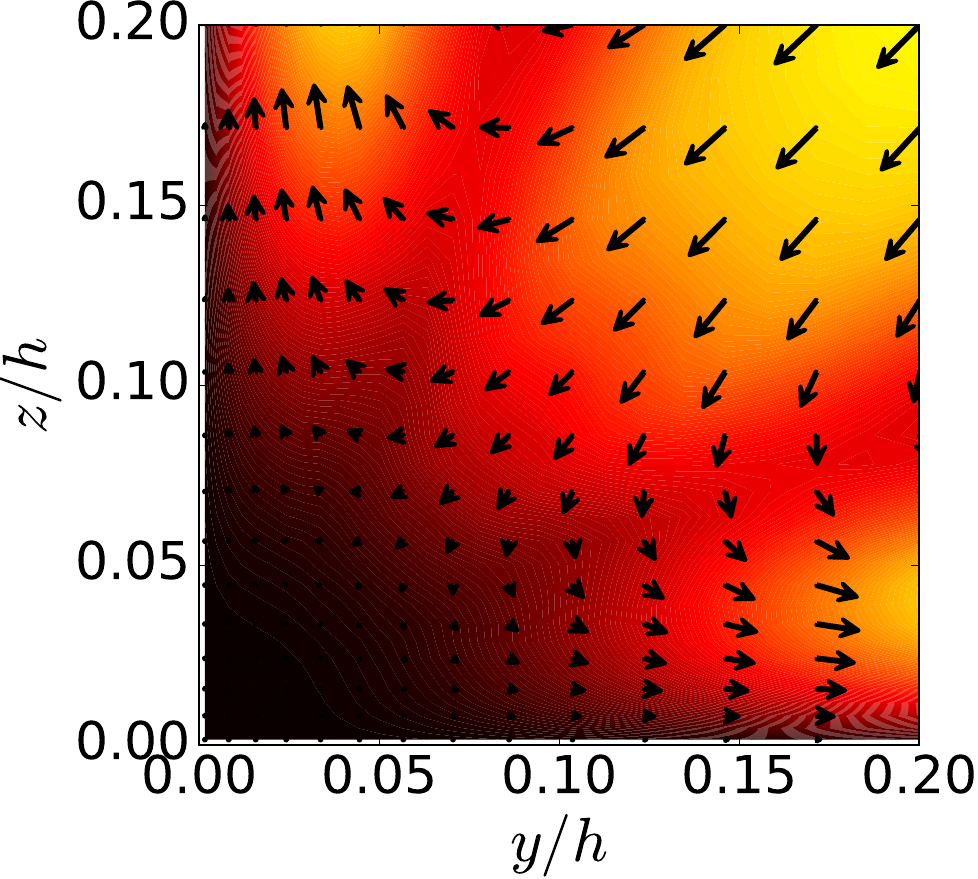}}	 
	\caption{Contour and vector plots of secondary mean motions of (a) baseline (b) DNS
		and (c) machine-learning-predicted results at the corner region. The color of contour 
		represents the magnitude of secondary velocity ($\sqrt{U_y^2 + U_z^2}$).}
	\label{fig:U_cont}
\end{figure}

The mean secondary velocity profiles for $U_y$ and $U_z$ obtained by propagating the 
PIML-predicted Reynolds stress field are shown in Figs.~\ref{fig:U_f57}a and~\ref{fig:U_f57}b, respectively. 
To facilitate comparisons, the RSTM baseline and DNS results are also plotted in the same figures. 
The spatial variation of the RSTM simulated velocity basically captures the trend of the truth, 
but its magnitude is not predicted well. Notable deviations from the DNS results can be observed. 
Especially in the near-corner region, where the secondary motion is strong, the discrepancies are 
large in both RSTM simulated $U_y$ and $U_z$. In contrast, the PIML-predicted results show much 
better agreements with the DNS results. The improvements are even more pronounced in the 
regions where secondary flow is strong (e.g., $y/h = 0.25$ and $0.5$). This can be seen clearly in 
the contour plot of the secondary flow by zooming in the near-corner region (Figure~\ref{fig:U_cont}). 
In this region the mean flow pattern simulated by RSTM is notably different from the DNS results. 
For the RSTM modeled mean secondary motion (Fig.~\ref{fig:U_cont}(a)), the flow approaches to the 
corner along the diagonal, and its velocity only decreases very close to the corner ($y/h< 0.05$ 
and $z/h<0.05$).
However, the DNS data show a different flow pattern, where the magnitude of the flow velocity
towards the corner decreases earlier. The secondary velocity is significantly reduced after 
$y/h < 0.15$ and $z/h < 0.15$. Its magnitude decreases to almost zero as $y/h < 0.1$ 
and $z/h < 0.1$. Comparing Figs.~\ref{fig:U_cont}b and~\ref{fig:U_cont}c, the propagated 
flow field from PIML prediction excellently captures the general pattern of the DNS results 
in the near-corner region. It shows a significant improvement over the baseline results,
suggesting that PIML-predicted Reynolds stresses are superior and can provide a better 
in-plane velocity field, especially in the region with a strong secondary flow. Slight discrepancies 
still exist in the region with a mild secondary flow. For example, PIML-predicted $U_z$ profile 
at $y/h = 1$ deviates from the DNS results (Fig.~\ref{fig:U_f57}b). A possible reason is that 
the training data may contain small errors introduced in the interpolation process, which can 
cause notable velocity discrepancies in regions where the secondary flow is weak. Similar 
discrepancies can be found in the velocity profiles propagated with the DNS Reynolds stresses, 
which are shown in Fig.~\ref{fig:U_DNS}b. 

\subsubsection{Merits of Expansion of Invariant Feature Space}
One of the novelties in this work lies in applying an integrity invariant basis on the construction 
of input feature space in the PIML learning-prediction process. Compared to the set of ten input 
features used in~\cite{Wang2016}, the current feature space is markedly expanded. A pertinent 
question is what benefits the expanded feature space offers? Or in other words, would it be possible 
to achieve a similar success by using an incomplete invariant basis as the input space, e.g., the ten 
features used in~\cite{Wang2016}. 
\begin{figure}[htpb]
	\centering 
	\includegraphics[width=0.4\textwidth]{u_f57_legend}\\
	\subfloat[$U_y$]{\includegraphics[width=0.45\textwidth]{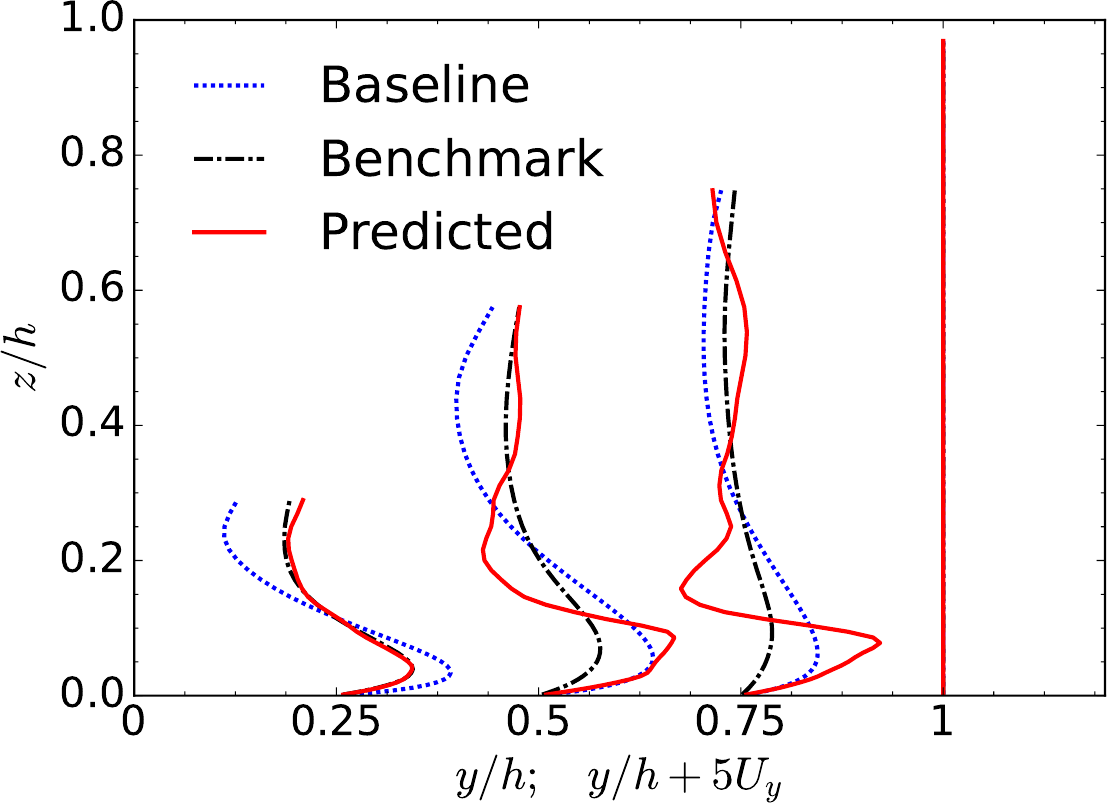}} 
	\subfloat[$U_z$]{\includegraphics[width=0.45\textwidth]{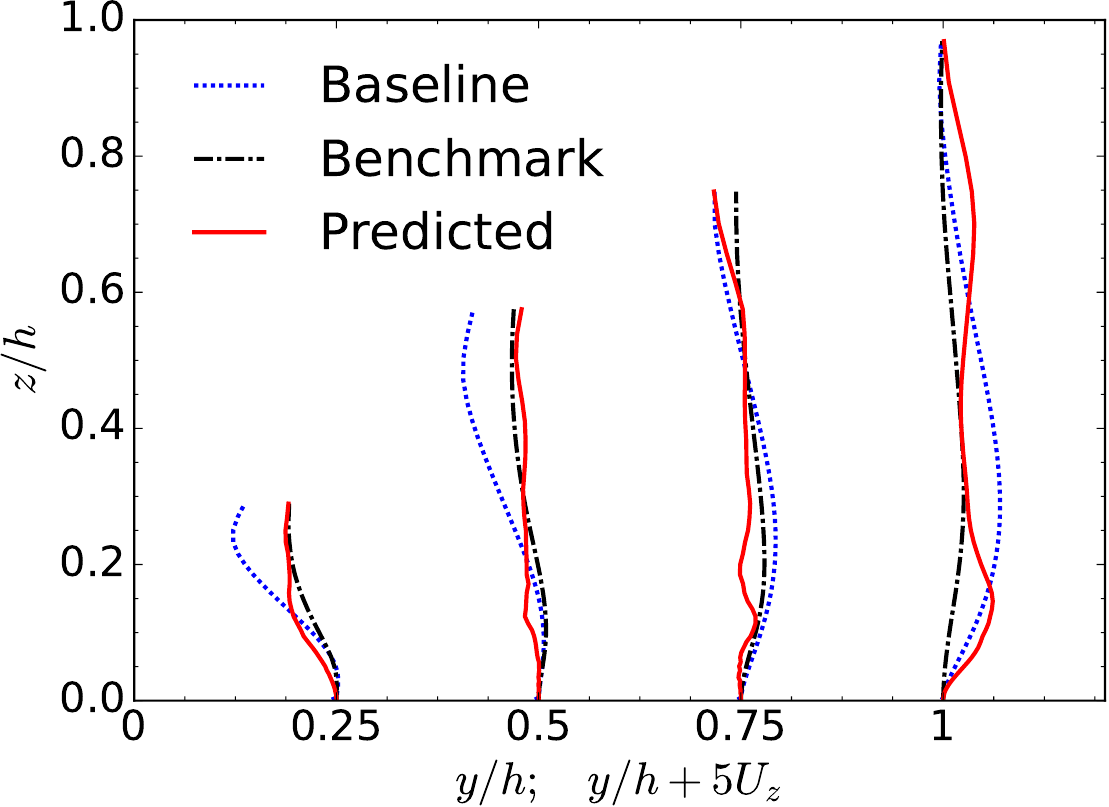}}
	\caption{In-plane velocity profiles (a) $U_y$ and (b) $U_z$ obtained by 
		propagating the PIML-predicted Reynolds stresses via RANS equations 
		at Reynolds number $Re = 3500$. The baseline (RSTM) and DNS results are also plotted for comparison. The discrepancy functions of
		Reynolds stresses are trained on ten features used in~\cite{Wang2016}.}
	\label{fig:U_f10}
\end{figure}
To investigate this issue, we perform the same training, prediction, and propagation steps
for the same test case shown above, but with the input set of ten features used in \cite{Wang2016} 
instead of current 57 features. The propagated mean secondary velocity profiles $U_y$ and $U_z$ 
from the corrected Reynolds stress field are shown in Figs.~\ref{fig:U_f10}a and~\ref{fig:U_f10}b, 
respectively. The profiles of PIML-predicted velocity are improved over the baseline RSTM results in 
regions close to the corner ($y/h = 0.25$) for both components $U_y$ and $U_z$. However, in the 
regions away from the corner ($y/h = 0.5$ to $0.75$), the profiles of machine-learning predictions 
largely deviate from the DNS results and become even worse than the baseline predictions. Unphysical 
wiggling of the velocity profiles can be observed in both figures. Compared to the mean velocity 
with the expanded feature space shown in Fig.~\ref{fig:U_f57}, the accuracy of propagated mean 
motion with the input set of ten features significantly deteriorates. 

\begin{figure}[htbp]
	\centering
	\includegraphics[width=0.4\textwidth]{u_f57_legend}\\
	\subfloat[normal stress imbalance]{\includegraphics[width=0.45\textwidth]{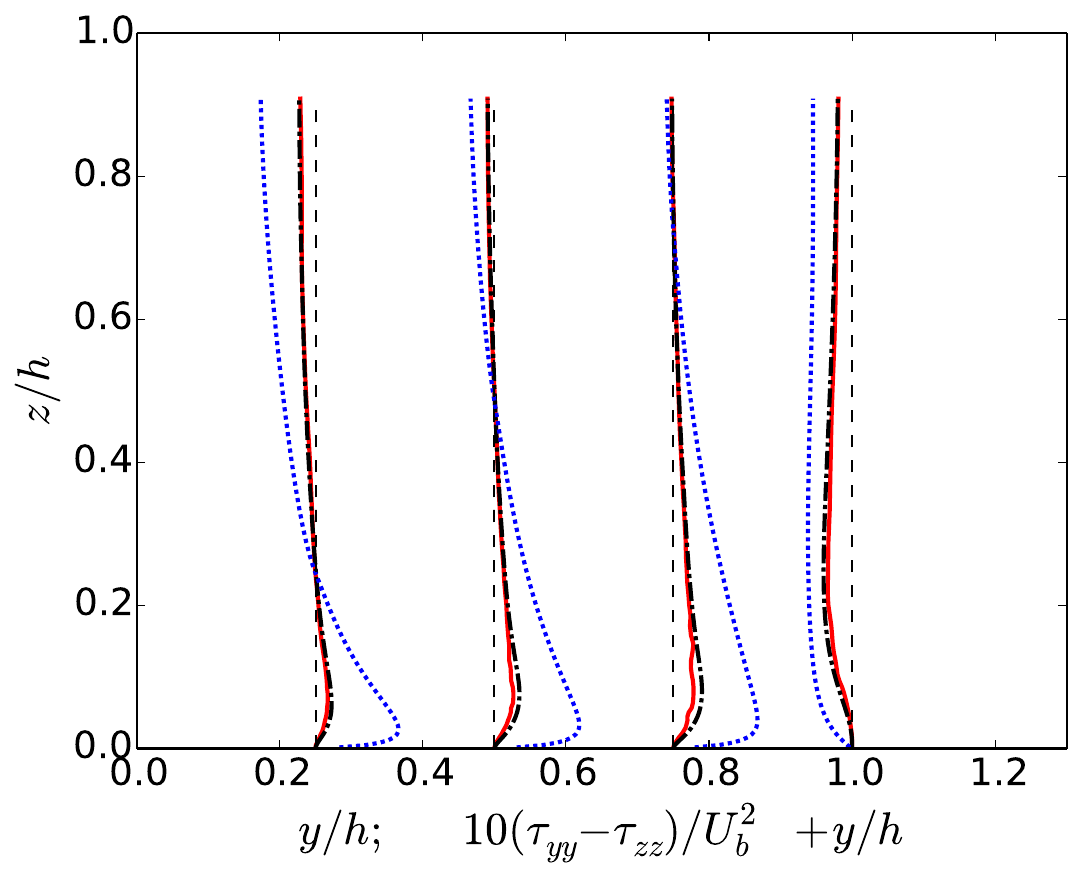}}
	\subfloat[shear component]{\includegraphics[width=0.45\textwidth]{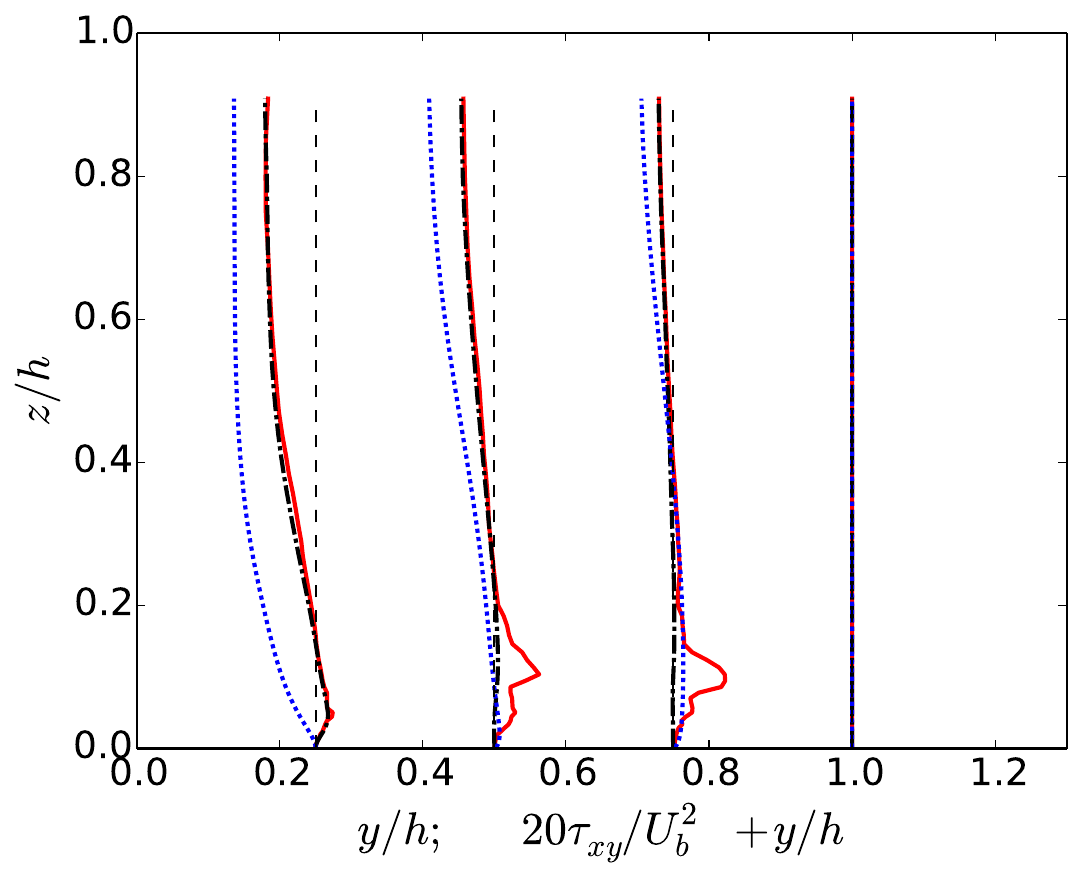}}
	\caption{Profiles of (a) normal stress imbalance $\tau_{yy} - \tau_{zz}$ and (b) shear component $\tau_{xy}$ of 
		corrected Reynolds with the discrepancy model trained on 10 features used in~\cite{Wang2016}. 
		The profiles are shown at four streamwise locations $x/H = 0.25, 0.5, 0.75,$ and $1$. 
		Corresponding DNS and baseline (RSTM) results are also plotted for comparison.}
	\label{fig:tau_10}
\end{figure}
The deterioration of the mean flow prediction indicates that the PIML-corrected Reynolds stresses 
are not satisfactory to propagate to an improved mean velocity field. Figures~\ref{fig:tau_10}a and~\ref{fig:tau_10}b show profiles of the normal stress imbalance term ($\tau_{yy} - \tau_{zz}$) 
and shear stress component $\tau_{xy}$ of the corrected Reynolds stresses with the input space 
of ten features. It can be seen that the machine learning predictions are significantly improved over 
the baseline RSTM results since the profiles of both terms have a better agreement with the DNS 
results. The norms of discrepancies between the prediction and the truth are significantly reduced 
compared to those of the RSTM results. However, a notable difference from the machine learning 
predictions with 57 features (Fig.~\ref{fig:tau}) is that the profiles of the predicted Reynolds stress 
with ten input features wiggle at the lower part of the duct. Especially for the shear 
component~$\tau_{xy}$, several bumps can be clearly found on the profiles at $y/h = 0.5$ and 
$0.75$. Although the overall Reynolds stress predictions are improved (i.e., discrepancies in 
tensor components are reduced), the derivative of the turbulent shear stress field becomes even worse 
than the baseline in these regions with wiggles and bumps. These unphysical wiggles can pollute
the propagated velocity field since it is the divergence of Reynolds stress appearing in 
the momentum equation and determining the velocity propagation. This explains the significant 
deterioration of the velocity prediction in Fig.~\ref{fig:U_f10}a.

The expanded input space based on an integrity invariant basis used in this work significantly 
improves the learning and prediction performances of Reynolds stress discrepancies, and thus 
improved mean velocity predictions can be obtained through the RANS propagation. The 
merits of applying the expanded input set to construct the random forest model are twofold. 
First, the predicted field from the random forest model tends to be non-smooth due to its 
pointwise estimation. The level of non-smoothness increases when the dimension of the input 
space is lower than the truth. In other words, if the features are not rich enough to differentiate 
different response points in feature space, the prediction tends to be more non-smooth due to 
projecting onto an incomplete basis. Using a set of complete invariant bases significantly 
increases the ``resolution'' of input space. Thus, the prediction 
performance with expanded features (Fig.~\ref{fig:tau}) is markedly superior over that with the 
previous input set of ten features (Fig.~\ref{fig:tau_10}). Note that the increase in resolution of 
input space is more important for learning discrepancies in the orientation of the Reynolds stress 
tensor, since eigenvectors contain more information than the eigenvalues do, and thus more features 
are needed to explain them. Second, the current 57 input features also include rotational invariants other 
than full invariants, which also can improve the learning performance of angle discrepancies, since 
the Euler angles are not reflection invariant. To improve the capability of generalization, 
one possible approach would be to expand the training data to include reflected states of the system and then 
to teach the model to be fully invariant on this expanded training set~\cite{ling2016machine}. 
On the other hand, we can try to explore better parameterizations of eigenvector system instead of using Euler 
angles. To improve the representation of discrepancies in the eigenvectors is an ongoing research, 
which is out of the scope of the current work. 

It is worth noting that using the expanded feature basis increases the risk of overfitting because it introduces more free parameters into the model.  In this study, the model performance was evaluated on a single flow configuration at multiple Reynolds numbers.  It is very likely that performance on other flow configurations would be unreliable due to overfitting.  Further validation is necessary to assess the generalization of these models.  Nevertheless, it is already useful to have a model that is specific to a given flow configuration because it is very common in industry to run many simulations of closely related flows. 

\section{Discussion}
\label{sec:discussion}
\subsection{Concept of ``physics-informed machine learning''}
The concept of ``physics-informed machine learning'' claimed in this work is to emphasize the 
attempt to consider and embed physical domain knowledge into every stage of the machine 
learning process, including the construction of the input feature space, choice of output 
responses, and learning-prediction process. For most physical problems, data are not rich 
enough to conduct the traditional machine learning, since most of existing algorithms have 
been developed for business applications and have rarely been used for physical systems. 
Ling et al.~\cite{ling2016machine} demonstrated the merits of embedding physical truth
(e.g., invariance properties) into the machine learning process. Arguably, we believe that the 
state-of-the-art machine learning techniques have difficulties in learning the hard constraints 
in a physical system (e.g., conservation law, realizability) from any reasonable amount of data.
Therefore, in the proposed method the machine learning is employed to correct the RANS
model instead of replacing it. Moreover, physical hard constraints (e.g., realizability of Reynolds 
stress) and domain knowledge (e.g., reasoning for choosing raw features) are incorporated. 
We emphasize the concept of physics-informed machine learning to draw attentions from 
the audiences in both physical modeling and machine learning communities. We try to 
demonstrate that data-driven modeling is a promising compensate for traditional physical 
modeling. At the same time, we have incorporated as much turbulence domain knowledge 
as possible instead of entirely depending on data. 

\subsection{Challenges and Perspectives of the Current Framework}
Although it has been demonstrated that the current PIML framework is a promising way to
improve predictive turbulence modeling, there are a few of challenges associated with the
propagation of corrected Reynolds stresses to the mean velocity field, which need improvement
in future work. Here we briefly discuss these challenges. In both~\cite{Wang2016} and current 
work, the RANS-modeled Reynolds stresses are shown to be significantly improved. However, to 
propagate the success in Reynolds stress predictions to QoIs (e.g., mean velocity field) is still 
challenging. First, we should acknowledge that the improvements in Reynolds stresses are from 
the point view of pointwise estimation. It is possible that the predictions are close to the truth but 
are not smooth (i.e., slightly wiggling around the truth), which might pollute the propagated 
velocity field. This is because the currently used machine learning algorithm, i.e., random forest, 
may not necessarily improve the spatial derivative of Reynolds stress field due to the pointwise 
statistics.  Second, the numerical stability could be another issue that affects a robust propagation. 
The second issue is relatively trivial and can be solved by using some numerical tricks, e.g., adding
artificial diffusion terms. The first issue on the non-smoothness of machine learning predictions
is the main roadblock for velocity propagation. How to effectively use the information of the spatial
correlation of the Reynolds stress field from the data is crucial to further improve the current
framework. One possible method would be to assume a non-stationary spatial correlation 
structure, whose hyper-parameters can be determined based on data and physical prior knowledge.
The pointwise machine learning predictions could be regulated by this correlation structure to 
ensure the smoothness physically. Finally, it is worth noting that in this paper we have yet to 
demonstrate the ability to generalize the predictive performance more broadly for a wide range
of flows, since only one flow configuration at multiple Reynolds numbers is considered. 
To generalize the predictive capability by using more comprehensive databases with various flow physics will be the subject of future study.

\section{Conclusion}
\label{sec:conclusion}
Recently, the growing availability of high-fidelity data sets has led to increased interest in using
data-driven approaches to improve the predictive capability of RANS models. 
Wang et al.~\cite{Wang2016} demonstrated that the RANS-modeled Reynolds stresses can 
be improved by learning the functional form of the Reynolds stress discrepancy from available 
data. However, it is still an \emph{a priori} study, since whether these improved Reynolds stresses 
can be propagated to obtain a better velocity field remains unclear. In this work, we 
introduce and demonstrate the procedures toward a complete Physics-Informed Machine
Learning (PIML) framework for predictive turbulence modeling, including learning the Reynolds stress 
discrepancy function, predicting Reynolds stresses in different flows, and propagating the 
predicted Reynolds stresses to the mean flow field. To improve the learning-prediction performance, 
the input features are expanded by constructing an integrity invariants basis of given
raw mean flow variables. The predictive accuracy of the velocity field by propagating the PIML-corrected 
Reynolds stresses is investigated. The fully developed turbulent flow in a square duct 
is used as the test case. The discrepancy functions are trained in the flows at low Reynolds 
numbers and used to predict a different flow at a high Reynolds number. The numerical results show 
excellent predictive performances in both Reynolds stresses and their propagated velocity field, 
demonstrating the merits of the proposed PIML approach in predictive turbulence modeling. 

\section*{Acknowledgment}
Sandia National Laboratories is a multi-mission laboratory managed and operated by Sandia Corporation, 
a wholly owned subsidiary of Lockheed Martin Corporation, for the U.S. Department of Energy’s National 
Nuclear Security Administration under contract DE-AC04-94AL85000. SAND2017-0664 J

\section*{Compliance with Ethical Standards}
Conflict of Interest: The authors declare that they have no conflict of interest.







\end{document}